\documentclass[12pt,nofootinbib,eqsecnum,longbibliography,prd]{revtex4-2}

\usepackage{amsmath,amssymb}
\usepackage[american]{babel}
\usepackage[export]{adjustbox}
\usepackage{microtype}
\usepackage{color}
\usepackage{graphicx}
\usepackage{hyperref}
\hypersetup{colorlinks, citecolor=blue, linkcolor=black, urlcolor=blue}
\usepackage{bm}
\usepackage{tikz-feynman} 
\usepackage{anyfontsize}
\usepackage{dsfont}
\usepackage{soul}

\newcommand{\ket}[1]{|#1\rangle}
\newcommand{\bra}[1]{\langle #1|}

\newcommand{\del}{\partial}

\newcommand{\newvar}{b}

\definecolor{purple}{rgb}{0.4,0.0,0.4}

\begin{document}

\title{Probing ``Continuous Spin'' QED with Rare Atomic Transitions}

\author{Aidan Reilly}
\email{areilly8@stanford.edu}
\affiliation{SLAC National Accelerator Laboratory, 2575 Sand Hill Road, Menlo Park, CA 94025, USA}
\author{Philip Schuster}
\email{schuster@slac.stanford.edu}
\affiliation{SLAC National Accelerator Laboratory, 2575 Sand Hill Road, Menlo Park, CA 94025, USA}
\author{Natalia Toro}
\email{ntoro@slac.stanford.edu}
\affiliation{SLAC National Accelerator Laboratory, 2575 Sand Hill Road, Menlo Park, CA 94025, USA}

\date{\today}

\begin{abstract}
An intriguing and elementary possibility is that familiar massless particles like the photon could be ``continuous spin'' particles (CSP) with a small but non-zero spin Casimir $\rho$. In this case, the familiar two polarization states of the photon are accompanied by an infinite tower of integer spaced helicity modes, with couplings dictated entirely by Lorentz symmetry and the parameter $\rho$. 
We present a formalism for computing bound state atomic transitions for scalar QED when $\rho \neq 0$, employing path integral methods not often used for bound state computations, but that readily generalize to the CSP case. 
We compute several illustrative amplitudes and show that $\rho\neq 0$ opens new decay channels for atomic transitions with rates controlled by $\rho\alpha/\omega$ for transition frequency $\omega$. These new channels can appreciably modify the rates of ``forbidden'' transitions. For example, the lifetime of the hydrogen $2s$ state would be affected at $O(1)$ for $\rho\sim 0.1$ eV, suggesting new directions for fundamental tests of QED in laboratory experiments. 
\end{abstract}

\maketitle
\newpage
\tableofcontents


\newpage
\section{Introduction}
\label{sec: Introduction}
Wigner famously classified all massless particles in 3+1 dimensions, showing that they can be described by irreducible representations of the ISO(2) Little Group \cite{Wigner:1939cj}. Additional restrictions on their interactions come from the covariance of soft amplitudes \cite{Weinberg1, Weinberg2, Weinberg3} and consistency of gauge and general relativity couplings \cite{WeinbergWitten, Berends:1984rq, Berends:1984wp,Bekaert:2010hp}, which imply that the only long range forces possible are those mediated by massless particles with spin 0,1, or 2 (i.e. scalar particles, vector gauge theories, or general relativity). However, these arguments rely on the assumption that massless particles have helicity eigenstates that transform trivially under boosts. This ignores the most general representations of the ISO(2) Little Group, which are defined, for any four momentum $k^\mu$, by their spin Casimir $W^2 = -\rho^2$, where 
\begin{equation}
W^\mu = \frac{1}{2}\epsilon^{\mu\nu\rho\sigma}k_\nu J_{\rho\sigma},
\end{equation}is the Pauli-Lubanski vector and $J_{\rho\sigma}$ is the covariant generator of rotations in 3 dimensions. We call $W^2$ the spin Casimir because for massive particles, $W^2 = -m^2S(S+1)$, where $m$ is the mass of the particle and $S$ is its spin. For massless particles, $\rho$, which we call the spin scale (with units of momentum), parameterizes the degree of helicity mixing under Lorentz boosts, just as spin does in the massive case. Familiar theories enforce $\rho=0$, and hence assume that helicity is a Lorentz invariant quantity. 
The lack of theoretical development of the $\rho\neq 0$ alternative has limited our ability to even pose meaningful questions about its compatibility with nature. 

A particle with nonzero $\rho$, called a continuous spin particle (CSP), necessarily requires a description with an infinite tower of helicity modes. 
In particular, a particle in the state $\ket{k,h}$, with four-momentum $k$ and integer helicity $h$, will mix with every other integer helicty eigenstate under an infinitesimal Lorentz boost $\Lambda$ in a transverse direction:
\begin{equation}
    U(v_\perp)\ket{k,h} = \sum_{h'}c_{hh'}\ket{\Lambda k,h'},
\end{equation}
with $h'\in \mathbb{Z}$. The coefficients $c_{hh'}$ are determined by the ISO(2) Little Group representation \cite{Schuster:2013pxj}. For $\rho\neq 0$, all $c_{hh'}$ are non-zero but in the high-energy limit $k^0 \gg \rho$ they approach $\delta_{hh'}$, with helicity-mixing $c$'s suppressed by powers of $\rho v_\perp/k^0$. 

Because a CSP involves states with arbitrarily high helicity eigenvalues, a description of even a free CSP resisted traditional Lagrangian field theory treatments up until the work of \cite{Schuster:2014hca}. Moreover, the classic results constraining high spin particles derived from soft limits of scattering amplitudes were generalized to the CSP case only recently in \cite{Schuster:2013pxj,Schuster:2013vpr,Schuster:2014hca}. What was uncovered was surprising -- non-trivial soft limits of CSP-matter scattering amplitudes are compatible with unitarity and Lorentz invariance, and in the $\rho\rightarrow 0$ limit, CSPs faithfully recover the known soft scattering amplitude results with {\bf only} the $h=0,1,2$ modes interacting in this limit. Thus, it would appear that (some or all) known theories of massless particles could be limits of CSP theories, and perhaps known forces could be mediated by CSPs with small $\rho$ \footnote{CSPs were previously thought to have infinite heat capacity due to the infinite number of polarization modes. However, \cite{Schuster:2024wjc} recently showed that despite the infinite number of degrees of freedom, CSPs have effectively finite heat capacity because the tower of helicity modes takes infinite time to thermalize, and in particular only one $\pm h$ mode thermalizes efficiently for $\rho\lesssim T$.}. 

Building on the field theory description of CSPs in \cite{Schuster:2014hca}, \cite{Schuster:2023xqa} studied CSP dynamics and derived CSP interactions with spinless matter particles described in the worldline formalism. Like the soft emission limits in \cite{Schuster:2013vpr}, the CSP interactions studied in \cite{Schuster:2023xqa} were found to be either scalar-, vector-, or tensor-like, meaning that in the $\rho\to 0$ limit and at energies $\gg \rho$ they reduce to familiar massless scalar, vector, or tensor field interactions, respectively.  Furthermore, deviations from known formulas (e.g. the Larmor frequency) were derived for a photon with non-zero spin scale (CSP photon). These deviations were shown to be controlled by powers of $\rho v/\omega$, with $\omega$ the photon's energy and $v$ the matter particle's speed, again echoing the scaling of the soft limit theorems.  These results suggest that a small but non-zero $\rho$ leads to familiar UV physics, with new IR deviations.  Taking this analysis one step further, \cite{Schuster:2023jgc} calculated free particle scalar QED amplitudes mediated by a CSP photon by extending the classical matter coupling in \cite{Schuster:2023xqa} to a quantum worldline formalism.  The result of that analysis showed how the probability of measuring a final state helicity other than 1 in elementary processes such as Compton scattering can be expanded order by order in $\rho$, with larger helicity deviations requiring large powers of $\rho$. Interestingly, while all known photon observations are consistent with $\rho=0$, they are also consistent with a non-zero $\rho  \lesssim$ neV, with the strongest constraints on the size of $\rho$ coming from estimated (though not robust) effects on stellar cooling. It is therefore worthwhile to explore precision laboratory tests of $\rho$.

With the requisite tools now in place, the observation that a CSP photon can carry away large values of angular momentum, with a probability proportional to powers of $\rho$, leads us to the present work. From the perspective of atomic transitions, this feature has the direct consequence that single photon emissions traditionally forbidden by angular momentum conservation are now possible. One would expect that transitions of this type are also suppressed for energy splittings $\gg \rho$, by powers of $\rho$ over kinematic invariants, with higher powers being required for transitions more highly forbidden. We will see that this is precisely the case. One can therefore set limits on the size of $\rho$, or potentially even observe a non-zero $\rho$, by searching for photons emitted via such forbidden transitions. We must be careful when carrying out this analysis, however, as the state-of-the-art only includes spin-0 matter particles coupled to a bosonic CSP, not a fermionic spin-1/2 matter particle. We therefore must contend with forbidden transitions that are relativistically suppressed but in principle allowed, e.g. via spin-flip interactions (so called magnetic dipole transitions). We will not specifically address this issue here, as the purpose of this paper is to set up the techniques required to perform bound state transition amplitudes mediated by CSP photons. Nonetheless, conservative bounds can always be set even ignoring magnetic dipole interactions (and other higher order QED terms). 

The rest of the paper will be set up as follows: in Section \ref{Sec:pathIntegral} we develop a formalism to calculate bound state transition amplitudes via path integral techniques, first in regular QED and then generalizing to non-zero $\rho$. The main result of this section, and in some sense, the main result of this paper, is that in the non-relativistic limit the matrix element for a bound state atomic transition mediated by a CSP photon with helicity $h$ takes the simple form:
\begin{equation}
\begin{split}
    M_h &= (-q_e\sqrt{2}\tfrac{\omega}{\rho})\int_0^{2\pi} d\phi \ e^{ih\phi} \int d^3x\; \psi_{in}(\vec{x})\psi^*_{out}(\vec{x}-\tfrac{\rho}{\omega m}\vec{e}_\phi),
\end{split}
\end{equation}
where $q_e$ is the bound particle's charge, and $\vec{e}_\phi=\frac{1}{\sqrt{2}}(e^{i\phi}\vec{\epsilon}_-+e^{-i\phi}\vec{\epsilon}_+)$ is built from two linearly independent (unit norm) polarization vectors orthogonal to the photon propagation direction $\vec{k}$. As we will see explicitly in Section \ref{subsec: PI evaluation}, by Taylor-expanding the argument of the final-state wavefunction about $\vec x$ one recovers the familiar QED transition amplitude in the limit $\rho\rightarrow 0$. In Section \ref{sec: QHO} we apply these techniques to calculate transition amplitudes of a particle in a bound quantum harmonic oscillator (QHO) potential. The QHO serves as a simple toy model for which we derive amplitudes for a variety of forbidden transitions to highlight important features of CSP emissions. In Section \ref{sec: Hydrogen Atom} we repeat this analysis in the more physically relevant setting of a particle bound in a Coulomb potential (i.e. a hydrogen model), focusing on specific transitions of potential interest. The general analysis motivates a focus on the $2s\to 1s$ transition, from which we derive roough constraints on the spin-scale $\rho$ of the photon.  Finally, we conclude and discuss possible directions for future work in Section \ref{sec: conclusion}.

\section{Bound State Transition Amplitudes from the Path Integral}
\label{Sec:pathIntegral}
Non-relativistic transitions of the sort considered in this paper are typically derived from the matter Hamiltonian, treating the radiation field as a background.  
However, recent studies of CSP interactions with matter \cite{Schuster:2023xqa,Schuster:2023jgc} have worked exclusively in the Lagrangian formalism. Because the Lagrangian involves arbitrarily high powers of the matter particle's velocity, a closed-form Legendre transform to an equivalent \emph{interacting} matter Hamiltonian is not currently known in general. Therefore, our first step in analyzing transitions involving CSP photon emission will be to develop a method for computing bound state transition amplitudes starting from the matter Lagrangian, in the path integral formalism -- that is the purpose of this section\footnote{A complementary approach, which is equivalent for problems linear in the interaction, is to transform the interaction Lagrangian to \emph{free} Hamiltonian variables. That approach is illustrated in \cite{crossRefHyperfine}.}.

Constructing the path integral in a binding potential is straightforward, but the techniques to solve it exactly \cite{kleinert} are cumbersome.  It will be easier to take advantage of the simple evolution of energy eigenstates in the binding potential, essentially paralleling Hamiltonian perturbation theory.  
To do so, we first expand the path integral to linear order in the interaction Lagrangian describing the electron's coupling to the external CSP field.  The resulting path integral involves one insertion of the interaction Lagrangian at a time $t^*$ that must be integrated over, with paths weighted by the  static Lagrangian (kinetic + binding potential).  
We then break up the path-integral into three time-steps: from initial time $t$ to $t^*-\epsilon$, from $t^*-\epsilon$ to $t^*+\epsilon$, and from $t^*+\epsilon$ to the final observation time $t'$.  The initial and final path integrals do not involve the interaction, and are easily mapped to Hamiltonian evolution of the initial and final energy eigenstates.  By taking the intermediate interval, which includes the insertion of the interaction Lagrangian, to be infinitesimally small we can neglect the binding potential and use standard techniques for the free path integral. 

Note that we are focusing here on CSP corrections to the physics of transition radiation, while neglecting the corrections to the atomic binding potential and wavefunctions.  Because the transition radiation and binding potential in an atom arise from the same dynamics, a complete and consistent calculation must consider both.  However, this simplification is justified as an \emph{approximation} in precisely the transitions that are most physically interesting. Naively, both effects give rise to effects of $O(\rho^2 v^2/\omega^2)$ (where $\omega$ is the binding energy in the case of potential effects and the transition splitting in the case of radiation effects, and $v\sim\alpha$ is the characteristic speed of the bound electron).  However, our focus here is on forbidden transitions --- those whose width in QED is parametrically lower than naive dimensional analysis would suggest, because of angular momentum selection rules.  Angular momentum remains a good quantum number in the presence of non-zero $\rho$, and so CSP potential effects induce at most fractional corrections of order $O(\rho^2 v^2/\omega^2)$ to photon-mediated transitions. However, as we will see, changes to the radiation interaction open entirely new channels and therefore can alter a bound state's lifetime by $O(1)$ even for $\rho v/\omega \ll 1$, where the wavefunction modifications are still parametrically small. 
A similar hierarchy also arises in the case of fine-structure or hyperfine transitions, where $\omega_{radiation} \ll \omega_{binding}$ enhances the effects of non-zero $\rho$ in radiation.  Therefore, from here forward we take the binding potential to be purely a function of position and drop any CSP corrections to the potential (considering both a harmonic oscillator potential as a warm-up and a Coulombic potential for a more realistic model hydrogen atom).

\subsection{Hamiltonians and Lagrangians}
\label{subsec: hamiltonians and lagrangians}
We begin with a brief review of transition matrix elements in both the Hamiltonian and Lagrangian path-integral formalisms, introducing along the way some helpful notation.

We will assume that the Hamiltonian for the bound particle is given by 
\begin{equation}
 \begin{split}
    H &= H_0 + H_I, \\
    H_0 & = \frac{p^2}{2m} + V(\vec{x}),
    \end{split}\label{eq:Hexpansion}
\end{equation}
where $V(\vec{x})$ is the potential binding the particle, $H_I$ is the interaction term between the particle and the background photon field, and $p$ is the particle's momentum. For a standard ($\rho=0$)  photon, the interaction term is given by
\begin{equation}
    H_I = - \frac{q_e}{m} \vec{A}\cdot \vec{p},
\end{equation}
where $\vec{A}(\vec{x},t)$ is the photon field and $q_e$ is the particle's charge. The amplitude of moving from one bound eigenstate, $\psi_{\rm in}(t)$, to another, $\psi_{\rm out}(t')$, is thus given by
\begin{equation}
\label{eq: general amplitude}
    A = \bra{\psi_{\rm out}}e^{-i(H_0 + H_I)(t'-t)}\ket{\psi_{\rm in}} .
\end{equation}
In traditional perturbation theory, we can expand to first order in the interaction Hamiltonian to arrive at
\begin{equation}
    A \propto  \bra{\psi_{\rm out}}H_I\ket{\psi_{\rm in}}.
\end{equation}
Rather than take this route, however, we will transform Eq. \eqref{eq: general amplitude} into a path integral. We start by inserting two complete sets of position space basis vectors to yield
\begin{equation}
    \begin{split}
        \label{eq: A2}
    A = \int d^3x' \; \psi^*_{out}(\vec{x}',t') \int d^3x \; \psi_{in}(\vec{x},t)\; \bra{x'}e^{-i(H_0 + H_I)(t'-t)}\ket{x}
    \end{split}.
\end{equation}
We will call the last piece of this equation the propagator $K$:
\begin{equation}
    K(x',t',x,t) = \bra{x'}e^{-i(H_0 + H_I)(t'-t)}\ket{x}.
\end{equation}
We can then define the static propagator:
\begin{equation}
    K_0(x',t',x,t) = \bra{x'}e^{-iH_0(t'-t)}\ket{x},
\end{equation}
which gives the position space transition amplitude in the absence of coupling to external photons. 

The transition to the path integral formalism (and vice versa) is straightforward whenever the Hamiltonian is quadratic in momenta \cite{feynman}. In this case, repeatedly inserting the identity in terms of both position and momentum eigenstates at finely spaced times between $t$ and $t'$ generates a phase-space path integral over paths in $(p,x)$ space. The integrals over intermediate $p$ are Gaussian, and give rise to the position-space path integral form of the amplitude  Eq. (\ref{eq: A2}): 
\begin{equation}
    \begin{split}
        \label{eq: A PI}
        A &= \int d^3x'\; \psi^*_{out}(\vec{x}',t')\int d^3x \;\psi_{in}(\vec{x},t) \int_{\vec{z}(t)= \vec{x}}^{\vec{z}(t')=\vec{x}'} \mathcal{D}z \; e^{i\int_t^{t'} d\tau L(\vec{z},\tau,\dot{\vec{z}},\vec{A})},
    \end{split}
\end{equation}
where $L$ is given by the Legendre transform $L = \vec{p}\cdot \dot{\vec{x}} - H$ or equivalently, using \eqref{eq:Hexpansion},
\begin{equation}
    \begin{split}
    L &= L_0 + L_I,\\
    L_0 &=  \frac{1}{2}m\dot{\vec{z}}^2 - V(\vec{z}).
    \end{split}
\end{equation}

When the Lagrangian involves higher powers of $\dot x$, or the Hamiltonian higher powers of $p$,  subtleties arise in the equivalency above. The Legendre transform is single-valued for the CSP potential, but it is not analytically tractable. However, following \cite{Schuster:2023jgc}, we take Eq. \eqref{eq: A PI} using the interaction Lagrangian \cite{Schuster:2023xqa} as the starting point for computing CSP photon emission rates. This is sufficient for practical calculations. We leave further study of equivalent Hamiltonians to future work.  

\subsection{Splitting up the Path Integral}
\label{subsec: splitting the PI}
In the next two subsections we will derive transition amplitudes for ordinary QED, starting from the path integral above.  The approach generalizes straightforwardly to non-zero $\rho$, as we will discuss in Section \ref{subsec: Vhat CSP}.
We start by demonstrating how the path integral describing a bound state transition induced by coupling to a background field can be broken into three steps. 

For the special case of a background photon field, $L_I = q_e\dot{\vec{z}}\cdot\vec{A}(\vec{z},\tau)$. We can break this  field up into a sum of plane waves:
\begin{equation}
    \vec{A}(\vec{z},t) = \sum_j^N \vec{\varepsilon}_j e^{i\omega_{j} t - i\vec{k}_j\cdot \vec{z}},
\end{equation}
where $\vec{\varepsilon}$ is a polarization vector, and for single photon emission $N=1$. We then expand the part of the exponential involving the interaction with $\vec{A}$ as a power series, and keep only the leading order term. In this way we can pull down a vertex insertion:
\begin{equation}
\label{eq: Vhat QED}
    \int_t^{t'} d\tau\; \hat{V}_k^\varepsilon(\dot{\vec{z}},\tau) = (iq_e) \int_t^{t'} d\tau \;\vec{\varepsilon}\cdot\dot{\vec{z}} e^{i\omega \tau-i\vec{k}\cdot \vec{z} }.
\end{equation}
Note that while we have plugged in the Lagrangian interaction term $L_I$ for QED, this procedure works for generic $L_I$ that couples to a weak background field, and will generalize readily to the CSP case as we will see later. Plugging Eq. \eqref{eq: Vhat QED} into Eq. (\ref{eq: A PI}) gives us  
\begin{equation}
    \begin{split}
        A &= \int d^3x'\; \psi^*_{out}(\vec{x}',t')\int d^3x \;\psi_{in}(\vec{x},t) \int_{\vec{z}(t)= \vec{x}}^{\vec{z}(t')=\vec{x}'} \mathcal{D}z \; \left(\int_t^{t'} dt^* \hat{V}_k^\varepsilon(\dot{\vec{z}},t^*)\right) \;e^{i\int_t^{t'} d\tau L_0(\vec{z},\tau)},
    \end{split}
\end{equation}
where we have written the vertex insertion time variable as $t^*$ for clarity. By changing the order of the path integral and the integral over $dt^*$, we can notice three distinct time intervals of importance. At all times except for an infinitesimally small time period surrounding $t^*$, the non-interacting, bound state path integral is being evaluated. Therefore, in the time periods before and after $t^*$ we can replace the path integral with the static propagators $K_0(x_1,t^*-\epsilon,x,t)$ and $K_0(x',t',x_2, t^*+\epsilon)$. By utilizing the integral over initial and final state wave functions, one can see that this amounts to a simple phase evolution of each energy eigenstate. Thus, we are left with the only remaining path integral over the time $t^* \mp \epsilon$. The last observation we make here regards the fact that for the infinitesimal integral over $d\tau$, any terms not proportional to $\dot{z}$ (and hence $1/\epsilon$ in discretized form), will be negligible. This allows us to effectively drop the potential term from $L_0$ and write
\begin{equation}
    \lim_{\epsilon\rightarrow 0}\int \limits_{t^*-\epsilon}^{t^* + \epsilon} d\tau \; L_0 = \lim_{\epsilon\rightarrow 0}\int \limits_{t^*-\epsilon}^{t^* + \epsilon} d\tau \; \frac{1}{2}m\dot{\vec{z}}^{2}.
\end{equation}
One has to be careful about this step depending on the form of $V(\vec{z})$, especially for those potentials which appear singular as $\vec{z}\rightarrow 0$, as is the case for a Coulomb potential. Nonetheless, we verify that this step is valid in Appendix \ref{app: dropping potential} for common potentials that we will use later, including the Coulomb potential. Putting all of these observations together, we arrive at an amplitude that takes the form:
\begin{equation}
\label{eq: PI amp 1}
\begin{split}
        A &= e^{i\xi}\int_t^{t'} dt^* e^{-i(E_{in}-E_{out})t^*}\int d^3x_1 \; d^3x_2 \;\psi^*_{out}(\vec{x}_2)\;  \psi_{in}(\vec{x}_1)   \int \limits_{\vec{z}(t^*-\epsilon)= \vec{x}_1}^{\vec{z}(t^* + \epsilon)=\vec{x}_2}\mathcal{D}z \; \hat{V}_k^\varepsilon(\dot{\vec{z}},t^*)e^{i\int_{t^*-\epsilon}^{t*+\epsilon} d\tau \frac{1}{2}m\dot{\vec{z}}^2},
    \end{split}
\end{equation}
where $\xi = i(-E_{out}(t'-\epsilon) + E_{in}(t + \epsilon))$ is a simple phase factor. We can see from this final expression that the only remaining path integral is the free one with a single vertex insertion. Path integrals of this type are well studied in the literature \cite{Corradini:2015tik, Schuster:2023jgc, Strassler:1992zr}, and we employ many of the same techniques in the next section.

\subsection{Free Path Integral Evaluation}
\label{subsec: PI evaluation}
In order to evaluate the remaining path integral in Eq. \eqref{eq: PI amp 1}, we can make use of some string-inspired techniques. The trick is to realize that for sufficiently simple vertex operators $\hat{V}$, the path integral can be made into a Gaussian and become exactly solvable. If we can write $\hat{V}$ in terms of an exponential in phases linear (or quadratic) in $\vec{z}$ and $\dot{\vec{z}}$, then the path integral will be Gaussian. For the QED vertex operator used in Sec. \ref{subsec: splitting the PI} we can achieve this by writing 
\begin{equation}
\label{eq: V qed}   \hat{V}_{\rm seed}(\dot{\vec{z}},t) = q_e e^{i\omega t - i\vec{k}\cdot\vec{z} + i\vec{\varepsilon}\cdot\dot{\vec{z}}},
\end{equation}
 and defining $\hat{V}_k^\varepsilon(\dot{\vec{z}},t)$ as the piece of $\hat{V}_{\rm seed}$ linear in $\vec{\varepsilon}$. The main benefit here is that we can perform the full path integral calculation with $\hat{V}_{\rm seed}$, and then take only the piece linear in $\vec{\varepsilon}$ at the end of the calculation. We will see later that we can use the same $\hat{V}_{\rm seed}$ for a CSP photon insertion. In order to solve the path integral in closed form, it will be helpful to further put it into an unconstrained Gaussian form. Therefore, we want to integrate over all possible end points $\vec{x}_1$ and $\vec{x}_2$, independent from the wave functions $\psi_{in}$ and $\psi_{out}$. We can achieve such an integration naturally by Fourier transforming the wave functions so that the integrals over $\vec{x}_1$ and $\vec{x}_2$ apply only to end points of the path integral. We are now set up to solve the remaining path integral in Eq. \eqref{eq: PI amp 1} and find an amplitude that depends only on the specific wave functions in question. After making the change of variables $\tau \rightarrow \tau - t^*$, and putting $\hat{V}$ into exponential form, we find:
\begin{equation}
    \begin{split}
    \label{eq: amp after FT}
        A_{\rm seed} &= q_e \;e^{i\xi} \int dt^* e^{-i(E_{in}-E_{out}-\omega)t^*}\int \frac{d^3p_1}{(2\pi)^{3/2}} \frac{d^3p_2}{(2\pi)^{3/2}} \tilde{\psi}_{in}(\vec{p}_1)\tilde{\psi}^*_{out}(\vec{p}_2)\\
        &\int d^3x_1 d^3x_2\int\limits_{\vec{z}(-\epsilon) = \vec{x}_1}^{\vec{z}(\epsilon) = \vec{x}_2} \mathcal{D}z\; e^{\left[i\int\limits_{-\epsilon}^{\epsilon}d\tau \frac{m}{2}\dot{\vec{z}}(\tau)^2\right] - i\vec{k}\cdot \vec{z}(0) + i\vec{\varepsilon}\cdot \dot{\vec{z}}(0) + i\vec{p}_1\cdot \vec{z}(-\epsilon) - i\vec{p}_2\cdot \vec{z}(\epsilon)}.
    \end{split}
\end{equation}
In order to make this integral convergent we will make the mass re-definition $m = i\tilde{m}$. This is effectively the same as analytically continuing to Euclidean time by replacing $\tau \rightarrow -i\tau$ and $k_0 \rightarrow -ik_0$, but lets us keep track of fewer transformations. In particular, none of the terms in the exponential outside of the brackets will be affected by this transformation. An equivalent transformation is to send $m \rightarrow m +  i\tilde{m}$, and take $\tilde{m} \rightarrow 0$ at the end of the calculation. This path integral takes the same form as the unconstrained path integral in Eq. (3.6) of \cite{Schuster:2023jgc}, up to a factor of $2\tilde{m}$ and the exponentiated vectors being 3 dimensional rather than 4. Thus, we can use standard Green's function methods (see App. \ref{app: path Integral} for details) to find
\begin{equation}
    \begin{split}
        \label{eq: general amplitude momentum}
        A_{\rm seed} &= q_e e^{i\xi} \int dt^* e^{-i(E_{in}-E_{out}-\omega)t^*}\int d^3p_1 d^3p_2 \tilde{\psi}_{in}(\vec{p}_1)\tilde{\psi}^*_{out}(\vec{p}_2)\;e^{\frac{\vec{\varepsilon}}{2\tilde{m}}\cdot(\vec{p}_2 + \vec{p}_1)}\delta^{(3)}(P),
    \end{split}
\end{equation}
where $P = \vec{p_1} - \vec{p_2} - \vec{k}$. It will now be interesting to Fourier transform back to position space, as many wave functions are easier to work with in position space.  Taking the inverse Fourier transform and performing the integrals over $\vec{p}_1,\vec{p}_2,$ and $\vec{x}_2$ in Eq. \eqref{eq: general amplitude momentum}, we arrive at
\begin{equation}
\label{eq: general amp PI solved}
    \begin{split}
        A_{\rm seed} &= q_e e^{i\xi} \int dt^* e^{-i(E_{in}-E_{out}-\omega)t^*}\; e^{\frac{i\vec{\varepsilon}\cdot\vec{k}}{2m}} \int d^3x_1 \psi_{in}(\vec{x}_1)\psi^*_{out}(\vec{x}_1-\tfrac{\vec{\varepsilon}}{m}) e^{-i\vec{k}\cdot \vec{x}_1},
    \end{split}
\end{equation}
where we have re-introduced the real mass $m$. It is worth stopping here to consider how this seed amplitude relates to the full the QED solution. For a massless, transverse photon, $\vec{\varepsilon}\cdot \vec{k} =0$, and in the non-relativistic expansion $\vec{k}\cdot\vec{x}\ll 1$, simplifying Eq. \eqref{eq: general amp PI solved}. When we then take the term linear in $\vec{\varepsilon}$ to extract the QED amplitude from $A_{\rm seed}$, we find 
\begin{equation}
    A = (iq_e)  e^{i\xi} \;\vec{\varepsilon} \cdot \int dt^* e^{-i(E_{in}-E_{out}-\omega)t^*} \int d^3x_1 \psi_{in}(\vec{x}_1)(\tfrac{i}{m}\del_{\vec{x}_1})\psi^*_{out}(\vec{x}_1),
\end{equation}
which is just 
\begin{equation}
e^{i\xi} \int dt^* e^{-i(E_{in}-E_{out}-\omega)t^*}(-iq_e)\vec{\varepsilon} \cdot \bra{\psi_{out}}\frac{\vec{p}}{m}\ket{\psi_{in}},
\end{equation}
where we interpret $-i\del_{\vec{x}_1} = \vec{p}$. We have thus recovered familiar perturbation theory via a direct path integral calculation. It is also worth noting that when we finally square the amplitude to extract a physical observable, the remaining time dependent term will yield an energy conserving delta function. That is, in the limit of long observation time $(t'-t)$,
\begin{equation}
    \left| \int_t^{t'} dt^* e^{-i(E_{in}-E_{out} -\omega)t^*}\right|^2 = 2\pi \delta(E_{in} - E_{out} - \omega) (t'-t),
\end{equation}
but plays no other role in the calculation \cite{sakurai1967advanced}. We will therefore define the general seed amplitude, in the non-relativistic limit of a transverse photon, as  
\begin{equation}
\begin{split}
\label{eq: general result}
    A_{\rm seed} &= M_{\rm seed}\times e^{i\xi}\int dt^* e^{-i(E_{in}-E_{out}-\omega)t^*},\\
    M_{\rm seed} &= \mathcal{C} \int d^3x \psi_{in}(\vec{x})\psi^*_{out}(\vec{x}-\tfrac{\vec{b}}{m}),
\end{split}
\end{equation}
so that we can focus on the matrix element $M$. We have written $q_e\rightarrow \mathcal{C}$ and $\vec{\varepsilon}\rightarrow \vec{b}$ in order to keep the solution more general. We will see in the next section that Eq. \eqref{eq: general result} applies also to a CSP photon mediated transition with the appropriate substitutions for $\mathcal{C}$ and $\vec{b}$.

\subsection{Generalization to Non-zero $\rho$}
\label{subsec: Vhat CSP}
We now come back to the aforementioned observation that the derivation in the previous sections will hold true even in the case of a CSP photon emission, so long as we can write down a suitable Lagrangian interaction term $L_I$. An appropriate way to couple matter to a CSP photon that reduces to the regular photon coupling in the $\rho\rightarrow 0$ limit was derived in \cite{Schuster:2023xqa}, and used in a similar context to this paper for free particle Compton scattering in \cite{Schuster:2023jgc}. We refer the reader to Section II.b and II.c of \cite{Schuster:2023jgc} for a detailed overview of the operator, and we will simply quote that, in analogy with 
\eqref{eq: Vhat QED}, a vertex operator valid on and off-shell for a CSP photon takes the form:  
\begin{align}
V_{\gamma, CSP}^{k,\eta}(t) \equiv i q_e\, e^{ik\cdot z(t)} \ \left( e^{-i\rho\frac{\eta\cdot \dot{z(t)}}{k\cdot \dot{z}(t)}} \ \left(\sqrt{2} i k\cdot\dot{z}(t)/\rho \right) +{\cal D} X(\eta,\dot{z(t)}) \right),
\label{eq:offshell_vertex_operator}
\end{align}
where $\eta$ is an auxiliary space 4-vector used to parameterize different helicity modes of the CSP photon, and $\mathcal{D} = [k^2 - (-ik\cdot \eta - \frac{1}{2}(\eta^2 + 1)(-ik\cdot\del_\eta + \rho))]$ is the differential operator in the free CSP equation of motion for the Fourier modes, $\Psi(\eta,k)$, of the CSP gauge field. The function $X$ is not fixed by the arguments of \cite{Schuster:2023jgc}, but does not contribute to on-shell CSP amplitudes, so we can safely set $X=0$ without loss of generality. We therefore take the CSP vertex operator to be
\begin{equation}
\label{eq:CSP_vertex_operator_quotient_form}
\hat V_{CSP}^{k,\eta}(t) \equiv i q_e\, e^{ik\cdot z(t)} \ e^{-i\rho\frac{\eta\cdot \dot{z}(t)}{k\cdot \dot{z}(t)}} \ \left(\sqrt{2} i k\cdot\dot{z}(t)/\rho \right).
\end{equation}
In the fully non-relativistic limit, we can set $k\cdot z\approx \omega t$ and $k\cdot \dot{z} \approx \omega$. 
Using the additional fact that $\eta_0 =0$ (as we will see below), this approximation yields the non-relativistic CSP vertex 
\begin{equation}
    \label{eq: nonrel CSP vertex minimial}
    \begin{split}
    \hat{V}^{\rm non-rel}_{\text{CSP}}(t) &= (iq_e)  \;e^{i\omega t}e^{i\frac{\rho}{\omega}\vec{\eta}\cdot \dot{\vec{z}}(t)}\left(\sqrt{2}i\omega/\rho\right).
    \end{split}
\end{equation}
Immediately, we can notice that this operator is already in the form required for $\hat{V}_{\rm seed}$ of Eq. \eqref{eq: V qed}, after setting $\vec{k}\cdot \vec{z}\approx 0$. Therefore, to recover the non-relativistic CSP result from $\hat{M}_{\rm seed}$, we simply make the replacements $\vec{b} = (\rho/\omega)\vec{\eta}$ and $C = (iq_e)(\sqrt{2} i \omega/\rho)$, without needing to take only the linear part in $\vec{b}$. 

While we have therefore arrived at the conclusion that $A_{\rm seed}$ in Eq. \eqref{eq: general result} applies also a CSP photon emission, there is one more step we must take to get a physical result after making the appropriate $\mathcal{C}$ and $\vec{b}$ substitutions. In order to get an observable final state solution out of this operator, defined in the auxiliary $\eta$ space, we need to integrate it against the free CSP field Fourier amplitude $\psi_h$, over the appropriately regulated $\eta$ space measure defined in \cite{Schuster:2023xqa}. This process can be carried out for outgoing CSP photons using the following identity:
\begin{equation}
\label{eq: phi eta integrals}
\begin{split}
V_{CSP}^{out,k,h} &\equiv \int [{\bar d}^4\eta] \psi_h^*(\eta,k) \hat V_{CSP}^{k,\eta} = \int_0^{2\pi} \frac{d\phi_\eta}{2\pi} e^{-ih\phi_\eta} V_{CSP}^{k,\eta(\phi_\eta)},
\end{split}
\end{equation}
where $\psi_h$ is the wavefunction defined in \cite{Schuster:2023xqa}, and the $\eta$-integral expression can be simplified using identities in \cite{Schuster:2023xqa} to the form on the right-hand side with $\eta(\phi) \equiv (0,\vec{\eta}(\phi))$, and $\vec{\eta}(\phi)$ a reference polarization vector orthogonal to $\vec{k}$ with unit norm. This is not the only choice for $\eta(\phi)$ that one can take, but it is a convenient one for non-relativistic computations. In particular, we can take $\vec{\eta}(\phi)=\frac{1}{\sqrt{2}}(e^{i\phi}\vec{\epsilon}_-+e^{-i\phi}\vec{\epsilon}_+)$, built from two linearly independent (unit norm) polarization vectors orthogonal to the photon propagation direction $\vec{k}$. Different choices of vectors $\vec{\epsilon}_{\pm}$ will all give the same result. Each helicity vertex operator has the interpretation of a Fourier mode of $\hat V$, when $\eta$ is restricted to lie on a unit circle parameterized by $\phi$ \footnote{In some of the literature, it is this angular basis that is sometimes conflated with the term ``continuous spin", and has originated an unfortunate degree of confusion over the years. Helicity eigenvalues take on only discrete numbers, even for CSPs!}. 
The last step of every CSP photon emission calculation will therefore be to integrate over linear polarizations perpendicular to k with angular weight corresponding to helicity via the integral $\int_0^{2\pi}\frac{d\phi}{2\pi}e^{-ih\phi}$. 

We will end this section with a comment on the justification that we can drop all terms proportional to $\dot{\vec{z}}$ in the CSP photon vertex. One immediate objection may come from dropping terms proportional to $\vec{\dot{z}}/\rho$, which seem infinite in the $\rho\rightarrow 0$ limit. However, such terms are proportional to a total time derivative, and cancel by a version of the Ward identity that we prove in Appendix \ref{app: Ward}, as does the remaining $O(1/\rho)$ term in Eq. \eqref{eq: nonrel CSP vertex minimial}. It is still possible that the lowest order in $\rho$ vertex term will come with a non-zero power of $\vec{\dot{z}}$ that we may have erroneously dropped. However, we also detail in Appendix \ref{app: non-rel vertex} that for any spherically symmetric potential, the first potentially non-zero term occurs at $\mathcal{O}\left( (\frac{\rho v}{\omega})^n\right)$, for some value of $n$, determined by which transition is being analyzed. If this specific term were to be zero, then higher order velocity terms might be needed, but we do not find this to be the case in any of the scenarios analyzed. This justification does, however, rely on being able to apply the form of $A_{\rm seed}$ in Eq. \eqref{eq: general amp PI solved} to the CSP case, before taking any approximations. Thankfully, it was shown in shown in \cite{Schuster:2023jgc} that the full CSP vertex of Eq. \eqref{eq:CSP_vertex_operator_quotient_form} can be put into exponential form via a clever integration (see App. \ref{app: non-rel vertex}), so we are safe to use Eq. \eqref{eq: nonrel CSP vertex minimial}.

\section{3-D Quantum Harmonic Oscillator}
\label{sec: QHO}
In this section we will apply the techniques developed in Sec. \ref{Sec:pathIntegral} to an easy-to-solve, toy problem, namely a particle bound in a 3-D harmonic oscillator potential. Such potentials are ubiquitous in physics and will serve as a good warm up that will highlight many important features of bound state CSP photon emissions before moving on to a hydrogen atom model in Sec. \ref{sec: Hydrogen Atom}. The 3-D isotropic harmonic oscillator has a central potential that depends only on distance:
\begin{equation}
    V(x,y,z) = \frac{1}{2}m\Omega^2(x^2 + y^2 + z^2) = \frac{1}{2}m\Omega^2 r^2,
\end{equation}
(we use $\Omega$  for the quantum harmonic oscillator (QHO) frequency to distinguish it from the photon energy $\omega$). The QHO potential yields wavefunction eigenstates in spherical coordinates: 
\begin{equation}
    \psi_{q\ell m}(r,\theta, \phi) = N_{q\ell}r^\ell e^{-\nu r^2/2}L_q^{(\ell + \frac{1}{2})}(\nu r^2)Y_{\ell m}(\theta, \phi),
\end{equation}
where 
\begin{equation}
    \begin{split}
        &N_{q\ell} = \sqrt{\sqrt{\frac{\nu^3}{4\pi}} \frac{2^{q +\ell+3} q! \nu^\ell}{(2q+2\ell+1)!!}} \quad \text{is a normalization constant};\\
        &\nu = m\Omega;\\
        &L_q^{(\ell + \frac{1}{2})}(\nu r^2) \quad \text{is the generalized Laguerre polynomial,}\\
        &Y_\ell^m(\theta,\phi) \quad \text{is a spherical harmonic},
    \end{split}
\end{equation}
with $\ell$ and $m$ labeling the total angular momentum and its $\hat{z}$ component as usual, and $q$ counts the number of nodes in the radial wavefunction. The energy levels depend only on the principal quantum number $n = 2q + \ell$ as:
\begin{equation}
    E_n =  \Omega (n + \tfrac{3}{2}).
\end{equation} 

\subsection{$(q,\ell,m) : (0,\ell,\ell) \rightarrow (0,0,0)$ Transitions}
The simplest types of transitions we can calculate for this potential will be those for which the radial part of the wave function does not change between the initial and final state. We will call these types of transitions angular momenta only transitions. If we furthermore restrict ourselves to $m=\ell$ in the initial state, then the only difference in wave function dependencies between different $\ell$ states will come from $\mathcal{R}^\ell \sin^\ell\theta e^{i\ell \phi}$ in spherical coordinates, which becomes $r^\ell e^{i\ell\phi}$ in cylindrical coordinates, with r now the cylindrical radius. Cylindrical coordinates are ultimately the most natural for this problem as the spherical symmetry of the system is broken by the photon emission, but cylindrical symmetry remains. Finally, we also restrict our final state to always be the ground state, for simplicity. It will be useful to keep in mind that we can define a direction of propagation for the emitted photon (CSP or otherwise), which we will set to be along $\hat{z}$. This is not a completely general choice, as we have already chosen the angular momentum of the initial state to be quantized along the same direction. Nonetheless, a photon emitted in the $\hat{z}$ direction will dominate the phase space of the emission, and can be used to get a parametric understanding of the allowed processes. Let us now recall that for the generic bound state amplitude given in Eq. \eqref{eq: general result}, in the non-relativistic limit, the matrix element $M_{\rm seed}$ takes the form:
\begin{equation}
    M_{\rm seed} =  \mathcal{C} \int d^3x \; \psi_{in}(\vec{x})\psi^*_{out}(\vec{x}-\tfrac{\vec{b}}{m}).
\end{equation}
We can begin in Cartesian coordinates in order to address the dot product $\vec{b}\cdot \vec{x}$ that will come out of $\psi^*_{out}(\vec{x} - \frac{\vec{b}}{m})$. By only considering transitions to the ground state, the angular dependence on $\vec{b}$ will be the same for all values of $\ell$. Taking this first step we have:
\begin{equation}
\label{eq: Mseed start Dq=0}
    \begin{split}
        M_{\rm seed} 
        &= \mathcal{C}\tilde{N}_{00}\int d^3x\;\psi_{\rm in}(\vec{x}) \;e^{-\frac{\nu}{2}|\vec{x}|^2 -\frac{\nu}{2}|\frac{\vec{b}}{m}|^2 + \frac{\nu}{m}\vec{b}\cdot \vec{x}} ,
    \end{split}
\end{equation}
where $\tilde{N}_{q\ell}$ refers to the wave function normalization $N_{q\ell}$ multiplied by the normalization of the spherical harmonic $Y_{\ell,\ell}$, for example $\tilde{N}_{00} = \frac{1}{\sqrt{4\pi}}N_{00}$. 
Recalling that for a QED photon, $\vec{b}$ will eventually be replaced with the polarization vector $\vec{\varepsilon}$, while for a CSP photon, $\vec{b}$ is also proportional to the CSP polarization vector  $\vec{b} = \frac{\rho}{\omega}\vec{\eta}$, we will evaluate Eq. \eqref{eq: Mseed start Dq=0} for real vectors $\vec{b}$ that lie in the $\hat{x}$-$\hat{y}$ plane. Such an evaluation will apply to linearly polarized QED photons, or any CSP photon. For the circularly polarized QED photon that we will be interested in, we simply combine the results from $x$ and $y$ polarizations appropriately. 
We therefore write the dot product $\vec{b}\cdot\vec{x}$ in cylindrical coordinates as
\begin{equation}
\label{eq: b dot x CSP}
    \vec{b}\cdot \vec{x} =|\vec{b}|r\cos(\phi - \phi_b),
\end{equation}
where $\phi_b$ is the position in the $\hat{x}$-$\hat{y}$ plane of $\vec{b}$ . We can re-write $\phi_b = \phi -\tilde{\phi}$, sending the integration over angles $\phi$ to
\begin{equation}
    \int_0^{2\pi} d\phi \rightarrow \int_{-\phi_b}^{2\pi - \phi_b}d\tilde{\phi}.
\end{equation}
Because the function being integrated is periodic in $\tilde{\phi}$, we can then simply shift the integration to $0\rightarrow 2\pi$. Ultimately this trick will pull out an overall $e^{i\ell \phi_b}$ and simplify the integration over angles. Using Eq. \eqref{eq: b dot x CSP} and plugging in $\psi_{\rm in}(x) = \psi_{0,\ell,\ell}$, we find 
\begin{equation}
    \begin{split}
        M_{\rm seed} &= \mathcal{C}\tilde{N}_{00}\tilde{N}_{0\ell}\;e^{i\ell\phi_b}\;e^{-\frac{\nu|\vec{b}|^2}{2m^2}}\;\int drdz d\tilde{\phi} \; r\;r^\ell\;e^{-\nu(r^2 + z^2) -iBr\cos\tilde{\phi} + i\ell\tilde{\phi}},
    \end{split}
\end{equation}
where $ B = \frac{i\nu}{m}|\vec{b}|$. For the the QED case, we set $\phi_b = 0$ for a photon linearly polarized in the $\hat{x}$ direction, and $\phi_b = \pi/2$ for the $\hat{y}$ direction, while for a CSP photon $\phi_b$ becomes $\phi_\eta$. The integral over $dz$ is simply a Gaussian, and the integral over $d\tilde{\phi}$ yields a function involving an $\ell$'th order Bessel function, which we can then integrate against the remaining $r$ dependencies to yield 
\begin{equation}
\label{eq: QHO angular momenta M general}
    \begin{split}
        M_{\rm seed} = \mathcal{C}\tilde{N}_{00}\tilde{N}_{0\ell}\pi^{3/2}\;e^{i\ell\phi_b}e^{-\frac{\nu|\vec{b}|^2}{4m^2}} \;  \frac{|\vec{b}|^\ell}{2^{\ell}\nu^{3/2}m^\ell}.
    \end{split}
\end{equation}
To go further and get a physically understandable amplitude, we now need to decide whether to extract the QED or CSP photon solution, which we do in the following sections. 
\subsubsection{Extracting the QED Result}

We can now notice from the form of Eq. \eqref{eq: QHO angular momenta M general} that the only transition with a non-zero piece linear in $\vec{b}$ comes from $\ell=1$. Each piece of the polarization vector will therefore yield a component (when setting $\mathcal{C} = q_e$, and linearizing in $\vec{b} = \vec{\varepsilon}_{x,y}$):
\begin{equation}
    \begin{split}
        M_{x,y}(\phi_b) &= e^{i\phi_b}q_e\left[ \frac{ 1}{2\nu^{3/2}m }\right]\; \tilde{N}_{00}\tilde{N}_{01}\;\pi^{3/2} .
    \end{split}
\end{equation}
Finally, by setting $\phi_b = 0$ for $M_x$, $\phi_b = \pi/2$ for $M_y$, and combining the two to get a circularly polarized photon with angular momentum quantized along the $\hat{z}$ axis, we get a total amplitude:
\begin{equation}
\label{eq: QED QHO amp}
    M = \frac{1}{\sqrt{2}}(M_x -iM_y) = q_e\sqrt{\frac{\Omega}{2m}},
\end{equation}
where we have also plugged in the appropriate normalizations. One can easily verify that this is the exact matrix element that a perturbative Hamiltonian calculation yields, up to a normalization factor in the photon field $\vec{A}$. 
\subsubsection{Extracting the CSP Result}

To get the CSP result is actually somewhat simpler, as we can simply insert the proper definitions of $\vec{b} = (\rho/\omega)\vec{\eta}$ and $C = (iq_e)(\sqrt{2} i \omega /\rho)$ to get:
\begin{equation}
    M(\phi_\eta) = (iq_e)(\sqrt{2} i \omega /\rho)\left[ \tilde{N}_{00}\tilde{N}_{0\ell}\;\pi^{3/2}\;e^{i\ell\phi_\eta}e^{-\frac{\nu \rho^2 }{4\omega^2m^2}}\;  \frac{ (\rho/\omega)^\ell}{2^{\ell}\nu^{3/2}m^\ell }\right],
\end{equation}
where we have used the fact that $|\vec{\eta}|=1$. From this expression we can immediately see both the helicity support in $e^{i\ell\phi_\eta}$ and the power of $\rho$ dependence in $\rho^\ell$. To formally extract the helicity support, we need to integrate over linear CSP polarizations orthogonal to $\vec{k}$ as per Eq. \eqref{eq: phi eta integrals}. We therefore have that the matrix element for a given helicity mode is
\begin{equation}
\begin{split}
    M_h &= \int_0^{2\pi} \frac{d \phi_\eta}{2\pi}e^{-ih\phi_\eta} M(\phi_\eta).
\end{split}
\end{equation}
The integral
\begin{equation}
    \int_0^{2\pi} \frac{d \phi_\eta}{2\pi}e^{-i\phi_\eta(h - \ell)},
\end{equation}
will yield 1 for $h = \ell$ and 0 otherwise. Therefore, we have that the only non-zero helicity mode allowed by this transition is $h=\ell$, giving us 
\begin{equation}
\label{eq: A_l CSP}
    M_\ell =(iq_e)|P_\ell^\ell|\sqrt{\frac{2(2\ell+1)}{2^\ell(2\ell)!(2\ell+1)!!}}\;\  \left[\frac{\rho^{\ell-1}}{ \omega^{\ell-1}}\left(\sqrt{\frac{\Omega}{m}}\right)^{\ell}e^{- \frac{\nu\rho^2}{4\omega^2m^2}}\right],
\end{equation}
 where $|P_\ell^\ell|$ is the constant multiplying $\sin^\ell\theta$ in the associated Legendre polynomial $P^\ell_\ell(\sin\theta)$. Furthermore, in the limit that $\rho\rightarrow 0$, the only nonzero transition occurs for $\ell=1$, yielding a matrix element
\begin{equation}
    M_{\ell=1} = i q_e\sqrt{\frac{\Omega}{2m}}.
\end{equation}
This expression exactly matches up with the QED result in Eq. \eqref{eq: QED QHO amp}, up to an overall phase. For larger values of initial angular momentum $\ell$, Eq. \eqref{eq: A_l CSP} tells us that higher powers of $\frac{\rho v}{\omega}$ are required, where we have used the fact that $v\sim \sqrt{\Omega/m}$ for the QHO. This fact encapsulates the general result of this paper, that forbidden transitions can be mediated by a CSP photon at the price of a $\frac{\rho v}{\omega}$ suppression. While we have only considered transitions between the ground state and a state with no radial mode excitation here, conservation of angular momentum tells us that this pattern will persist for non-zero $n$ as well. In the following section, we will consider a different type of transition for the QHO, and see that it follows the same trend.

\subsection{A $\Delta \ell = 0$ Transition}
\label{subsec: QHO other transitions}
The previous section discussed generic transitions of the type $\Delta \ell = \Delta m = \Delta n$. While those transitions highlight important features of a CSP photon mediated emission, and can be computed rather generally, they do not encompass the only type of forbidden transition that a CSP photon can mediate. We will perform here another example of a transition forbidden by QED that differs from the angular momenta only type, namely a $\Delta \ell = 0$ transition. Unlike the case of angular momenta only transitions, a $\Delta \ell = 0$ transition necessarily requires starting in an excited state of the radial part of the wave function. The simplest such transition, and the one that we will consider here, is the transition from $\psi_{1,0,0} \rightarrow \psi_{0,0,0}$, i.e. the first excited radial state down to the ground state. It is worth noting that due to the specific symmetry of the QHO, there is no $n=1, \ell=0$ state, so $\psi_{1,0,0}$ is actually an $n=2$ state. By angular momentum conservation, this transition requires emission of an $h=0$ CSP photon. Because $h=0$ is one integer away from the QED photon's $h=\pm 1$, we can expect that the lowest order contribution to this transition will come at first order in $\rho$. We now show that this is the case.

Starting again with our generic matrix element: 
\begin{equation}
    M_{\rm seed} = \mathcal{C}\; \int d^3x \; \psi_{in}(\vec{x})\psi^*_{out}(\vec{x}-\frac{\vec{b}}{m}),
\end{equation}
we can make the same initial steps as with the angular momenta only transitions because once again $\psi_{out}$ is the ground state. We can therefore write

\begin{equation}
    \begin{split}
        M_{\rm seed} = \mathcal{C}\tilde{N}_{00}\tilde{N}_{10}\;e^{-\frac{\nu|\vec{b}|^2}{2m^2}}\;\int drdz d\tilde{\phi} \; r\left(\frac{3}{2}-\nu(r^2 + z^2)\right)\;e^{-\nu(r^2 + z^2) -iBr\cos\tilde{\phi}},
    \end{split}
\end{equation}
where $ B = \frac{i\nu}{m}|\vec{b}|$. Taking the integrals over $dz$ and $d\tilde{\phi}$ we get 
\begin{equation}
    \begin{split}
        M_{\rm seed} &= \mathcal{C}\tilde{N}_{00}\tilde{N}_{10}\;e^{-\frac{\nu|\vec{b}|^2}{2m^2}}\int dr \;2\pi J_0(Br)e^{-\nu r^2}\left[r\sqrt{\frac{\pi}{\nu}}  - r^3 \sqrt{\pi\nu}\right],
    \end{split}
\end{equation}
with $J_0(Br)$ the zero'th order Bessel function. While $J_0(Br)$ naively has an $O(1)$ term independent of $\rho$, one can see that such a term is proportional to a total $r$ derivative of $r^2e^{-\nu r^2}$, and so reduces to boundary terms which do not contribute. In fact, this term was guaranteed to cancel by the Ward identity argument in Appendix \ref{app: Ward}, since it is proportional to $\frac{1}{\rho}$ when we replace $\mathcal{C}$ with $(iq_e)(\sqrt{2}i\frac{\omega}{\rho})$. Carrying out the final $r$ integral and plugging this expression back in to our full amplitude we find:
\begin{equation}
\begin{split}
    M_{\rm seed} &= -\mathcal{C}\; \tilde{N}_{00}\tilde{N}_{10}\;e^{-\frac{\nu|\vec{b}|^2}{2m^2}}\;e^{\frac{\nu |\vec{b}|^2}{4m^2}}\left[\frac{\pi^{3/2}|\vec{b}|^2}{4m^2\sqrt{\nu}}\right].
\end{split}
\end{equation}
From here we can see immediately that there is no term linear in $\vec{b}$, so the QED solution is zero, as expected. Performing the steps to recover the non-relativistic CSP photon solution, we insert the proper definitions of $\vec{b} = (\rho/\omega)\vec{\eta}$ and $C = (iq_e)(\sqrt{2} i \omega /\rho)$ and normalizations to get
\begin{equation}
\begin{split}
    M(\phi_\eta) = \; \frac{1}{\sqrt{12}}\;e^{-\frac{\nu\rho^2}{4\omega^2m^2}}\;\left[\frac{\rho}{\omega}\frac{\Omega}{m}\right].
\end{split}
\end{equation}
When we take the final integral over $d\phi_\eta$, we find helicity support only for $h=0$, with a suppression of $\tfrac{\rho v}{\omega}$ relative to allowed QED transitions, as expected for a helicity mode one integer value away from $h=1$. Other $\Delta\ell = 0 $ transitions are possible for higher initial energy states, and one can verify the same helicity support and $\rho$ dependence.

\section{Scalar Hydrogen Atom}
\label{sec: Hydrogen Atom}
In this section we move beyond the simple QHO oscillator model and highlight the application of our techniques to a more realistic problem of interest, namely a hydrogen atom. In particular, we will be concerned only with a scalar hydrogen atom, i.e ignoring any spin dependent corrections. Understanding how to couple CSP photons to fermions is an important subject that we leave to future work. Nonetheless, we can still learn something about how forbidden transition amplitudes will change in the presence of a CSP photon, even ignoring spin. The 3D coulomb potential is given simply by
\begin{equation}
    V(x,y,z) = -\frac{q_e^2}{\sqrt{x^2 + y^2 + z^2}} = -\frac{q_e^2}{r}.
\end{equation}
Such a potential yields wavefunctions in spherical coordinates:
\begin{equation}
    \psi_{n \ell m}(r,\theta,\phi) = N_{n\ell}\;e^{-\nu/2}\;\nu^\ell \;L_{n-\ell -1}^{2\ell+1}(\nu) \;Y_\ell^m(\theta,\phi),
\end{equation}
where 
\begin{equation}
    \begin{split}
        &N_{n\ell} = \sqrt{\left(\frac{2}{na_0}\right)^3\frac{(n-\ell-1)!}{2n(n+\ell)!}} \quad \text{is a normalization constant};\\ 
        &\nu = \frac{2r}{na_0};\\
        &a_0 = \frac{4\pi}{mq_e^2} \quad \text{is the Bohr radius};\\
        &L_{n-\ell-1}^{2\ell +1} \quad \text{is the generalized Laguerre polynomial};\\
        &Y_\ell^m(\theta,\phi) \quad \text{is a spherical harmonic}.
    \end{split}
\end{equation}
The energy levels for the Hydrogen atom also depend only on the quantum number $n$:
\begin{equation}
    E_n = \frac{a_0 q_e^2}{2 n^2}.
\end{equation}

\subsection{$(n,\ell,m) : (\ell+1,\ell,\ell) \rightarrow (1,0,0)$ Transitions, Order by Order}
\label{subsec: hydrogen angular momentum transitions}
Just as we did for the harmonic oscillator, we will begin by computing angular momenta only transitions that end up in the ground state, where the initial state satisfies $\ell = m$.  In this section, we will forego notation to keep the solution general to a QED photon, and use the values for $\mathcal{C}$ and $\vec{b}$ appropriate for a CSP photon from the start. As we saw, one can always recover the QED solution by taking the $\rho\rightarrow 0$ limit anyway. With these substitutions, bound state transition elements take the form 
\begin{equation}
\label{eq: Mseed IVA}
    M_{\rm seed} = (iq_e)(\sqrt{2} i \omega/\rho)\; \int d^3x \; \psi_{in}(\vec{x})\psi^*_{out}(\vec{x}-\tfrac{\rho}{\omega m}\vec{\eta}).
\end{equation}
We could take the same initial steps for dealing with dot products that we took in the case of a harmonic oscillator, starting with:
\begin{equation}
    M_{\rm seed}  = (iq_e)(\sqrt{2} i \omega/\rho)\tilde{N}_{10}\int d^3x\;\psi_{\rm in}(\vec{x}) \;e^{-\frac{|\vec{x}-\frac{\rho}{\omega m}\vec{\eta}|}{a_0}},
\end{equation}
where
\begin{equation}
    \begin{split}
    |\vec{x}-\frac{\rho}{\omega m}\vec{\eta}|
    &= \sqrt{|\vec{x}|^2 + \frac{\rho^2}{\omega^2 m^2}|\vec{\eta}|^2-2 \frac{\rho}{\omega m}\vec{\eta}\cdot\vec{x}}\;.
    \end{split}
\end{equation}

However, we will not get a Gaussian integral over $z$ due to the square root in the exponential. This square root will prove generally problematic, so for the hydrogen atom it becomes beneficial to work order by order in $\rho/\omega$ from the start. We replace $\psi_{out}^*(\vec{x}- \frac{\rho}{\omega m}\vec{\eta})$ in Eq. \eqref{eq: Mseed IVA} with its Taylor expansion, getting an expression for $M_{\rm seed}$:
\begin{equation}
    M_{\rm seed}= (iq_e)(\sqrt{2} i \omega/\rho)\; \int d^3x\; \psi_{in}(\vec{x}) \left[1 -i \frac{\rho}{\omega m}\vec{\eta}\cdot \vec{p} - \frac{1}{2}\frac{\rho^2}{\omega^2 m^2}\vec{\eta}\cdot P_{ij} \cdot \vec{\eta} + \mathcal{O}((\frac{\rho}{\omega})^3)\right] \psi_{out}^*(\vec{x}),
\end{equation}
where $\vec{p} = -i\tfrac{\partial}{\partial \vec{x}}$ and $P_{ij} = -\tfrac{\del^2}{\del x_i \del x_j}$ are momentum operators. Immediately we can notice that the zero'th order term in the expansion will give 0, as it is simply proportional to the overlap of two orthogonal wavefunctions. This is expected from the Ward Identity as discussed Sec. \ref{subsec: Vhat CSP}.

The QED amplitude corresponds to the first order term of the expansion, which is independent of $\rho$. Having already extracted the piece of the exponential dependent on the photon polarization through perturbation theory, the problem is easier to solve in spherical, rather than cylindrical coordinates.  The $\mathcal{O}(\rho^0)$ piece of the matrix element, in spherical coordinates, is given by:
\begin{equation}
    \begin{split}
    M_{\rm QED} = (-q_e\sqrt{2} ) \left(\frac{2}{(\ell+1)a_0}\right)^\ell\frac{\tilde{N}_{(\ell+1)\ell} \tilde{N}_{10}}{ma_0}e^{i\ell \phi_\eta}\int drd\theta d\tilde{\phi}\;r^{\ell+2}\sin^{\ell+2}(\theta)e^{i\ell\tilde{\phi}}\cos(\tilde{\phi})e^{-r\frac{(\ell+2)}{(\ell + 1)a_0}},
    \end{split} 
\end{equation} 
where we have made the substitution $\tilde{\phi} = \phi - \phi_\eta$, with $\phi_\eta$ the angle that $\vec{\eta}$ takes on in the $x-y$ plane. The integral over $d\tilde{\phi}$ will yield $\pi$ if $\ell=1$ and 0 otherwise. As mentioned, this term corresponds to the QED piece, so the only allowed transition should be one which changes angular momentum by 1. After setting $\ell=1$ and taking the remaining integrals, we get:
\begin{equation}
    \begin{split}
        M_{\rm QED} = (-q_e\sqrt{2} )\;\alpha\; \frac{2^4}{3^{4}},
    \end{split}\label{eq:2pto1s}
\end{equation}
where $\alpha$ is the fine structure constant. This result is once again readily verified in a perturbative, Hamiltonian QED calculation. 

Finally, let us look at the second order term in the expansion, which corresponds to the first $\rho$ dependent transition amplitude. This time the only non-zero transition amplitude occurs when  $\ell = 2$, for which the $\tilde{\phi}$ integral yields $\pi/2$. For an initial state with $n\neq0$, there also exists an $\ell=0 \rightarrow \ell=0$ transition at this same order in $\frac{\rho}{\omega}$, but we do not consider that case here.  Plugging in $\ell=2$ and taking the integrals in spherical coordinates, we find:
\begin{equation}
    \begin{split}
        M_{\rho} &=(-q_e) \; \alpha \; \frac{3}{20} \left(\frac{\rho}{\omega}\alpha\right),
    \end{split}
\end{equation}
which is parametrically the same as the allowed QED transition, suppressed by a factor of $\tfrac{\rho v}{\omega}$ (with $v\sim \alpha$). So, we have seen that even for a hydrogen atom, forbidden transitions can be mediated by a single CSP photon, suppressed by the spin scale $\rho$. From symmetry of the angular momentum structure we can note that this pattern persists, such that higher angular momentum transitions occur at higher orders in $\tfrac{\rho v}{\omega}$, and new corrections to the $\Delta \ell =1$ transition occur at $\mathcal{O}((\tfrac{\rho v}{\omega})^2)$.  Of course, we could also calculate the forbidden transitions where $\Delta n \neq \Delta \ell$, and the total angular momentum change $\Delta \ell$ (for $\Delta \ell = \Delta m)$ will still set the scale of the $\rho$ dependence. For example, one can verify that a $\Delta \ell = \Delta\ m = 0$ transition  will come at $\mathcal{O}(\tfrac{\rho v}{\omega})$, as was shown for the QHO potential in Sec. \ref{subsec: QHO other transitions} transition. We set about calculating one such transition of particular importance in the next section.\\

\subsection{2s $\rightarrow$ 1s Transition}
\label{subsec: 2s->1s}
In this section we compute the $(n,\ell,m):(2,0,0)\rightarrow (1,0,0)$ hydrogen transition amplitude. This is not only an example of a forbidden $\Delta \ell = 0$ transition, similar to what was done for the quantum harmonic oscillator in Sec. \ref{subsec: QHO other transitions}, but it is also of particular experimental importance as it is forbidden to all orders in standard, non-relativistic QED. In fact, the leading order contribution to the decay lifetime of the 2s hydrogen state comes from two photon emission. On the other hand, as we will see here, a single CSP photon can mediate this transition. We once more start with the matrix element seed for a CSP photon:
\begin{equation}
    M_{\rm seed} = (iq_e)(\sqrt{2} i \omega/\rho)\; \int d^3x \; \psi_{in}(\vec{x})\psi^*_{out}(\vec{x}-\tfrac{\rho}{\omega m}\vec{\eta}) ,
\end{equation}
and take the Taylor expansion of $\psi^*_{out}$. We can use the knowledge that the first term in the expansion cancels for all transitions, and that the second gives the forbidden QED piece, to jump straight to the third term which is proportional to $\rho/\omega$. Plugging in $\psi_{in}$ and $\psi^*_{out}$ we have, in spherical coordinates, 
\begin{equation}
    M_{\rm seed} = (\tfrac{-q_e}{\sqrt{2}} \tfrac{\rho}{\omega}) \tilde{N}_{20}\tilde{N}_{10}\int drd\theta d\phi \;r^2 \sin(\theta) \left(2 - \frac{r}{a_0} \right)e^{-r/2a_0}\left(-\frac{\vec{\eta}}{m} \cdot P_{ij}\cdot \frac{\vec{\eta}}{m}\right)e^{-r/a_0},
\end{equation}
where 
\begin{equation}
    \left(-\frac{\vec{\eta}}{m} \cdot P_{ij}\cdot \frac{\vec{\eta}}{m}\right) =\Bigg[\left(\frac{\sin^2(\theta)\cos^2\tilde{\phi})}{m^2a_0^2}\right)  - \left(\frac{1}{m^2a_0r}\right) + \left(\frac{\sin^2(\theta)\cos^2(\tilde{\phi})}{m^2a_0r}\right)\Bigg],
\end{equation}
with $\tilde{\phi} = \phi - \phi_\eta$. Plugging this in and taking the required integrals yields:
\begin{equation}
\begin{split}
M_{\rm seed} 
&= q_e \alpha \frac{2^3}{3^4} \left(\frac{\rho}{\omega}\alpha\right).
\end{split}\label{eq:2sto1sM}
\end{equation}
We have thus confirmed that a single CSP photon can mediate a 2s $\rightarrow$ 1s transition!   The remaining step of integrating over $d\phi_\eta$ simply tells us that the CSP photon emitted will have helicity $h=0$.  Relative to the allowed $2p\to 1s$ transition rate \eqref{eq:2pto1s}, this transition amplitude is suppressed by a factor of $\frac{1}{2\sqrt 2}\frac{\rho\alpha}{\omega}$.  Since the characteristic speed of the bound electron in hydrogen is $\sim\alpha$, this is precisely the $\rho v/\omega$ suppression characteristic of CSP partner-mode couplings. Since the helicity of the emitted CSP photon matches the $\Delta \ell$ of the transition, its emission experiences no further suppression!

\subsection{Interpreting constraints on $\rho$}
It is worth commenting briefly on what these results mean for experiment. Given that we have considered only a scalar hydrogen atom, our results are only valid for electric transitions, where spin effects are not important. Moreover, for CSP effects to be observable their rates must be at least comparable to the leading QED process mediating a similar transition. In the first row of Table \ref{table: helicty support}, we show the lowest order in $\rho$ at which CSPs can mediate a given $\Delta \ell$ transition, without requiring an electron spin-flip. In contrast, the second row shows the largest non-zero contribution to the same transition, mediated by a QED photon. As we can see, $\Delta \ell \neq 1$ ``forbidden'' transitions are in fact allowed in QED, but they are meditated by higher magnetic or electric moment interactions and therefore suppressed by powers of $\alpha$ relative to their electric dipole counterparts. 

In most of these cases, we find that the $\rho \alpha/\omega$ suppression of the CSP-emission amplitude can only dominate over the $\alpha$ suppression of the higher-multipole QED transitions if $\rho \gtrsim \omega$, leading to a rather weak bound. Moreover, although for example direct $3d\to1s$ transitions are forbidden in the electric dipole, the $3d$ state is short-lived because it can decay through other dipole transitions, e.g. to  $2p$.  Thus, these high-$\ell$ states are not particlarly efficient probes of CSP effects. 

Of particular interest, however, is the 2s$\rightarrow $1s transition calculated in the previous section. Not only does the 2s state have a particularly long lifetime, any single photon emission is fully forbidden to all orders in QED, unless relativistic corrections to the wavefunction are included, which gives a large suppression $\propto \alpha^{11}$ \cite{Sucher:1978wq}. In fact, the primary decay channel for the 2s hydrogen atom in regular QED is via two photon emission. To convert the matrix element computed in Sec. \ref{subsec: 2s->1s} into a decay rate, we need to use Fermi's golden rule with the appropriate normalizations \cite{sakurai1967advanced}. 
Because none of the states involved carry angular dependence, the integral over angular phase space is trivial. Using $\Gamma = \frac{\omega}{2\pi\hbar}|M|^2$ and the matrix element \eqref{eq:2sto1sM}, we find
\begin{equation}
\tau_{2s}^{\rm CSP} \approx \left(\frac{\omega^2}{\rho^2}\right)1.6\times 10^{-4} \text{ s}.
\end{equation}
Comparing this value to the known lifetime of $0.12$ s \cite{Johnson:1972zza, Sucher:1978wq}, we can say with certainty that $\rho \lesssim \omega \sqrt{10^{-3}}$, or else the dominant decay channel for the 2s state of the hydrogen atom would be through a single CSP photon emission, and the observed lifetime would be shorter. Using the fact that $\omega \simeq 10.2$ eV, this sets the spin scale at $\rho \lesssim \mathcal{O}(0.1 \text{ eV})$. This bound can be strengthened with precision measurement and calculation of the $2s$ lifetime, improving as the square root of the relative precision. 

This relatively weak bound leads one to consider more highly forbidden and lower energy transitions. The 2s state of a helium atom with both electrons aligned, for example, cannot decay via a two-photon electric dipole transition, and thus has a much longer lifetime than that of hydrogen. However, any decay at all requires a spin flip of the excited electron. Similarly, the ground state of hydrogen can decay due to a hyperfine energy shift, giving the famous 21cm line, but requires a spin flip to do so. These potentially sensitive decay channels strongly motivated understanding how to couple CSP photons to particles with spin. The required formalism and phenomenology will be developed in forthcoming papers \cite{crossRefFermions} and \cite{crossRefHyperfine}, respectively. Other possible detection schemes include the use of nuclear decay and subsequent gamma cascades with missing energy, such as the experiment proposed in \cite{Benato:2018ijc}. The energy of emitted photons in these decays are typically much larger than those of low lying atomic transitions, but the sensitivity due to coincidence with other cascade photons allow for high precision measurements.

\begin{center}
    \begin{table}
    \begin{tabular}
    {|p{2.5cm}|p{3.3cm}|p{3.3cm}|p{3.3cm}|p{3.3cm}|}
     \hline
     & \hfil {$\Delta \ell = 0$} & \hfil {$\Delta \ell = 1$} &\hfil {$\Delta \ell =2$} &\hfil {$\Delta \ell =3$} \\
     \hline
     &&&&\\
     \hfil CSP & \hfil $\mathcal{O}(\rho)$ & \hfil  $\mathcal{O}(1)$ & \hfil $\mathcal{O}(\rho) $ & \hfil $\mathcal{O}(\rho^2)$\\
     &&&&\\
     \hline
     &&&&\\
    \hfil QED & \hfil $M_1$ & \hfil $E_1$ & \hfil $E_2$ & \hfil $E_3$\\
     &&&&\\
     \hline
    \end{tabular}
    \caption{Largest non-zero contribution for a given hydrogen atom transition amplitude. Each column corresponds to a given orbital angular momentum transition. Each row corresponds to either a CSP photon with scalar coupling (as used throughout the paper), or a full QED photon. The entry then gives the lowest order non-zero contribution to the transitions. For the CSP photon, we parameterize the contribution by orders in $\rho$, while for the QED photon we parameterize the contribution by the type of transition required ($E_1 =$ Electric Dipole, $M_1 =$ Magnetic Dipole, $E_2 = $ Electric Quadrupole, $E_3$ = Electric Octopole). It should be noted that $M1$ transitions can only mediate $\Delta \ell=0, \Delta n\neq 0$ transitions when relativistic corrections to the wave function are included.}
    \label{table: helicty support}
    \end{table}
\end{center}
\section{Discussion and Outlook}
\label{sec: conclusion}
In this work, we have derived a method for calculating atomic transition amplitudes in the case that the photon is a continuous spin particle. We first showed how calculate bound state transitions directly in velocity space via path integral techniques for regular QED. We then extended this method to a CSP photon using the matter coupling of a CSP field derived in \cite{Schuster:2023xqa}. We confirmed, in a new setting, that the physics of a CSP photon differs from the familiar QED photon in the deep IR, parameterized by the ratio of the CSP spin scale $\rho$ to its energy $\omega$, times the matter particle's velocity $v$. In particular, we have shown that forbidden atomic transitions can occur when mediated by a CSP photon, with rates suppressed by powers of $\tfrac{\rho \alpha}{\omega}$, proportional to how forbidden the transition is. After computing a few example transitions for a simple harmonic oscillator model, we calculated the matrix elements for various hydrogen atom transitions. Finally, we put these transition rates in context by comparing the spin scale suppression to the suppression coming from higher electric and magnetic moments that can mediate the same processes for regular QED. While moderate constraints on $\rho$ are set by analyzing the 2s$\rightarrow$ 1s hydrogen atom transition, further study is merited to understand the extent to which atomic transitions can be used to probe $\rho$. In addition to the study of spin flip transitions initiated in \cite{crossRefHyperfine}, studies of potential interest include the examination of more exotic atomic and nuclear transitions and of systems engineered to have longer-lived excited states.

\appendix

\section{Dropping the Potential Term in the Path Integral}
\label{app: dropping potential}
In this section we elaborate on the point made in Sec. \ref{subsec: splitting the PI}, that we may drop the potential term of the Lagrangian when computing the path integral over the infinitesimally short emission time. Let us start by recalling the argument made, that any contribution from the potential in the action becomes negligible: 
\begin{equation}
\begin{split}
    \lim_{\epsilon\rightarrow 0}\int \limits_{t^*-\epsilon}^{t^* + \epsilon} d\tau \; \frac{1}{2}m\dot{\vec{z}}^2 + V(\vec{z})  =\int \limits_{t^*-\epsilon}^{t^* + \epsilon} d\tau \; \frac{1}{2}m\dot{\vec{z}}^2.
\end{split}
\end{equation}
In order to understand why the velocity term remains while the potential term vanishes, we can look at the discretized path integral:
\begin{equation}
\begin{split}
     &\lim_{\epsilon \rightarrow 0}\int d^3x_1 d^3x_2 \psi_{in}(\vec{x}_1)\psi^*_{out}(\vec{x}_2) \int_{\vec{z}(t^*-\epsilon) = \vec{x}_1}^{\vec{z}(t^*+\epsilon) = \vec{x}_2} \mathcal{D}z\; \hat{V}(\dot{\vec{z}},t^*)\;e^{i\int_{t^*-\epsilon}^{t^*+\epsilon}d\tau L_0(\vec{z},\tau)}\\
     & = \lim_{\epsilon \rightarrow 0}\int d^3x_1 d^3x_2 \psi_{in}(\vec{x}_1)\psi^*_{out}(\vec{x}_2) \;\hat{V}(\tfrac{\vec{x}_2 - \vec{x}_1}{\epsilon},t^*) \;e^{i\epsilon\; L_0(\vec{x}_1,\vec{x}_2, \epsilon)} \\
     & = \lim_{\epsilon \rightarrow 0}\int d^3x_1 d^3x_2 \psi_{in}(\vec{x}_1)\psi^*_{out}(\vec{x}_2) \;\hat{V}(\tfrac{\vec{x}_2 - \vec{x}_1}{\epsilon},t^*) \;e^{i\; \epsilon\; \left[\frac{1}{2}m\left(\frac{\vec{x}_2-\vec{x}_1}{\epsilon}\right)^2 - V(\vec{x}_2)\right]}
\end{split}
\end{equation}
where we have made a couple of jumps. First, we have entirely gotten rid of the path integral measure $\mathcal{D}z$ by assuming $\vec{x}_1$ and $\vec{x}_2$ are a $\epsilon$ step away in the path integral. Due to this same description, instead of a factor of $2\epsilon$ that would show up naively, we have written only $\epsilon$, absorbing the factor of 2 into this small parameter. Futhermore, we have re-written velocity terms as $\underset{\epsilon\rightarrow 0}{\lim}\left(\tfrac{\vec{x}_2-\vec{x}_1}{\epsilon}\right)$. Writing velocity in this way highlights why it does not drop out of the exponentiated action term in the $\epsilon \rightarrow 0$ limit. The potential term on the other hand is well behaved in the $\epsilon \rightarrow 0$ limit so long as there is no regime in which $V(\vec{x}_2)\rightarrow \infty$. Strictly speaking this is not always true, but for the two potentials analyzed in this paper it is effectively true. For the Harmonic oscillator potential, the wave functions are Gaussian suppressed outside of a finite region, beyond which the amplitude has no support. We can therefore focus on a finite volume for $\vec{x}_2$, within which $\epsilon$ can be taken to be small enough that the potential term vanishes. The Coulomb potential, on the other hand, blows up as  $\vec{x}_2\rightarrow 0$, where the wavefunction does not vanish. However, there is a physical cutoff to the potential, given by the radius of the atom, $a_0$. We can therefore write the Coulomb potential as 
\begin{equation}
    V(x) = -\frac{q_1q_2}{|x| + a_0},
\end{equation}
so that $V(x)$ remains finite everywhere. If desired, $a_0$ can even be taken to 0 at the end of the calculation with  no issues. 

\section{Solving the Free Path Integral}
\label{app: path Integral}
In this section we solve the path integral from Sec. \ref{subsec: PI evaluation}, starting with Eq. \eqref{eq: amp after FT}:
\begin{equation}
    \begin{split}
        A &= \mathcal{C} \;e^{i\xi} \int dt^* e^{-i(E_{in}-E_{out}-\omega)t^*}\int \frac{d^3p_1}{(2\pi)^{d/2}} \frac{d^3p_2}{(2\pi)^{d/2}} \tilde{\psi}_{in}(\vec{p}_1)\tilde{\psi}^*_{out}(\vec{p}_2)\\
        &\int d^3x_1 d^3x_2\int\limits_{\vec{z}(-\epsilon) = \vec{x}_1}^{\vec{z}(\epsilon) = \vec{x}_2} \mathcal{D}z\; e^{\left[i\int\limits_{-\epsilon}^{\epsilon}d\tau \frac{m}{2}\dot{\vec{z}}(\tau)^2\right] - i\vec{k}\cdot \vec{z}(0) + i\vec{\varepsilon}\cdot \dot{\vec{z}}(0) + i\vec{p}_1\cdot \vec{z}(-\epsilon) - i\vec{p}_2\cdot \vec{z}(\epsilon)}.
    \end{split}
\end{equation}
In order to make this integral convergent we will make the mass re-definition $m = i\tilde{m}$.  It is also now useful to define the following variables:
\begin{equation}
    \begin{split}
    &j(\tau) = \frac{i}{2\tilde{m}}\left[\bar{j}(\tau) + \hat{j}(\tau)\right],\\
    &\bar{j}(\tau) = - \vec{p}_1 \delta(\tau + \epsilon) + \vec{p}_2 \delta(\tau - \epsilon) + \vec{k}\delta(\tau),\\
    &\hat{j}(\tau) = \vec{\varepsilon} \delta'(\tau).
    \end{split}
\end{equation}
Plugging in these substitutions our amplitude becomes:
\begin{equation}
    \begin{split}
    \label{eq: amp after j sub app version}
        A &= \mathcal{C} \;e^{i\xi} \int dt^* e^{-i(E_{in}-E_{out})t^*}\int \frac{d^3p_1}{(2\pi)^{d/2}} \frac{d^3p_2}{(2\pi)^{d/2}} \tilde{\psi}_{in}(\vec{p}_1)\tilde{\psi}^*_{out}(\vec{p}_2)\\
        &\times\int d^3x_1 d^3x_2\int\limits_{\vec{z}(-\epsilon) = \vec{x}_1}^{\vec{z}(\epsilon) = \vec{x}_2} \mathcal{D}z\; e^{-2\tilde{m} \int\limits_{-\epsilon}^{\epsilon}d\tau \dot{\vec{z}}(\tau)^2/4
        + j(\tau)\vec{z}(\tau)}.
    \end{split}
\end{equation}
This path integral takes the same form as the unconstrained path integral in Eq. (3.6) of \cite{Schuster:2023jgc}, up to a factor of $2\tilde{m}$ and the exponentiated vectors being 3 dimensional rather than 4. Thus, we can use all of the same tricks, first defining 
\begin{equation}
    \vec{w}(\tau) = \vec{z}(\tau) - \int\limits_{-\epsilon}^{\epsilon}d\tau' \; G(\tau,\tau')j(\tau'),
\end{equation}
where 
\begin{equation}
    \frac{1}{2}\del_\tau^2 G(\tau,\tau') = \delta(\tau-\tau'),
\end{equation}
(i.e. $G$ is the Green's function). It's also worth defining $\vec{w}_1$ and $\vec{w}_2$, which will simply be:
\begin{equation}
    \begin{split}
    &\vec{w}_{1} = \vec{w}(-\epsilon) = \vec{x}_1 - \int\limits_{-\epsilon}^{\epsilon}d\tau' \; G(-\epsilon,\tau')j(\tau'),\\
    &\vec{w}_{2} = \vec{w}(\epsilon) = \vec{x}_2 - \int\limits_{-\epsilon}^{\epsilon}d\tau' \; G(\epsilon,\tau')j(\tau').
    \end{split}
\end{equation}
From here, we can express $\vec{z}(\tau)$ in terms of $\vec{w}(\tau)$, $G$, and $j(\tau)$, and integrate by parts, yielding an expression that depends on $\vec{w}(\tau)$ only through boundary terms and the free action $\frac{\tilde{m}}{2}\dot{\vec{w}}(\tau)^2$. Thus we can write our amplitude as
\begin{equation}
    \begin{split}
        A &= \mathcal{C} \;e^{i\xi} \int dt^* e^{-i(E_{in}-E_{out})t^*}\int \frac{d^3p_1}{(2\pi)^{d/2}} \frac{d^3p_2}{(2\pi)^{d/2}} \tilde{\psi}_{in}(\vec{p}_1)\tilde{\psi}^*_{out}(\vec{p}_2)\\
        &e^{-2\tilde{m} \int\limits_{-\epsilon}^{\epsilon}d\tau d\tau' \frac{1}{2}j(\tau)G(\tau,\tau')j(\tau')}\\
        &\int d^3w_1 d^3w_2\int\limits_{\vec{w}(-\epsilon) = \vec{w}_1}^{\vec{w}(\epsilon) = \vec{w}_2} \mathcal{D}w\; e^{-2\tilde{m} \int\limits_{-\epsilon}^{\epsilon}d\tau \dot{\vec{w}}(\tau)^2/4
        }\\
        &\quad e^{-2\tilde{m}\left[\frac{1}{2}\int d\tau'\left(\vec{w}(\tau) +\frac{1}{2}\int d\tau'' G(\tau,\tau'')j(\tau'')\right)\del_\tau G(\tau,\tau')j(\tau')\right]_{\tau = -\epsilon}^{\tau = \epsilon}},
    \end{split}
\end{equation}
where we have used the fact that $\vec{w}$ is just a finite shift from $\vec{z}$ such that $\mathcal{D}z = \mathcal{D}w$, and that an integral over all $\vec{x}_{1,2}$ is the same as an integral over all $\vec{w}_{1,2}$. In the last two lines we have the massive, non-relativistic, free path integral and a boundary term, both of which will be integrated over $\vec{w}_1$ and $\vec{w}_2$. Starting with the remaining path integral, we have 
\begin{equation}
\begin{split}
    &\int\limits_{\vec{w}(-\epsilon) = \vec{w}_1}^{\vec{w}(\epsilon) = \vec{w}_2} \mathcal{D}w\; e^{-2\tilde{m} \int\limits_{-\epsilon}^{\epsilon}d\tau \dot{\vec{w}}(\tau)^2/4
        }= \left(\frac{\tilde{m}}{2\pi (2\epsilon)}\right)^{3/2}\;e^{-\frac{\tilde{m}(w_2-w_1)^2}{2(2\epsilon)} }.
\end{split}
\end{equation}
Now using the Green's function definition $G(\tau,\tau') = |\tau-\tau'|$, we can take the integrals over $w_1$ and $w_2$ to find: 
\begin{equation}
\begin{split}
\label{eq: final greens function integral}
\int d^3w_1 d^3w_2\left(\frac{\tilde{m}}{2\pi (2\epsilon)}\right)^{3/2}\;&e^{-\frac{\tilde{m}(w_2-w_1)^2}{2(2\epsilon)}-2\tilde{m}\left[\frac{1}{2}\int d\tau'\left(\vec{w}(\tau) +\frac{1}{2}\int d\tau'' G(\tau,\tau'')j(\tau'')\right)\del_\tau G(\tau,\tau')j(\tau')\right]_{\tau = -\epsilon}^{\tau = \epsilon} } \\
&= (2\pi)^{(3)}\delta^{(3)}(P),
\end{split}
\end{equation}
where $P = \vec{p}_1-\vec{p}_2-\vec{k}$. With that said, let us look at the integral
\begin{equation}
    -\tilde{m} \int\limits_{-\epsilon}^{\epsilon}d\tau d\tau'j(\tau)G(\tau,\tau')j(\tau') ,
\end{equation}
which has 3 pieces. First we can look at the $\bar{j} \circ \bar{j}$ piece:
\begin{equation}
    \begin{split}
    \bar{j}(\tau) \circ \bar{j}(\tau') &=\int d\tau d\tau' \bar{j}(\tau)G(\tau,\tau')\bar{j}(\tau) \\
    &= \int d\tau d\tau'\left[- \vec{p}_1 \delta(\tau + \epsilon) + \vec{p}_2\delta(\tau - \epsilon) + \vec{k}\delta(\tau)\right] |\tau - \tau'| \\
    &\quad \quad \times \left[- \vec{p}_1 \delta(\tau' + \epsilon) + \vec{p}_2\delta(\tau'- \epsilon) + \vec{k}\delta(\tau')\right] \\
    & = \int d\tau \left[- \vec{p}_1 \delta(\tau + \epsilon) + \vec{p}_2\delta(\tau - \epsilon) + \vec{k}\delta(\tau)\right] \times \left[- \vec{p}_1|\tau + \epsilon| + \vec{p}_2|\tau - \epsilon| + \vec{k}|\tau| \right] \\
    &= \vec{p}_1^2(0) -2\vec{p}_1\cdot \vec{p}_2(2\epsilon) - 2\vec{p}_1\cdot \vec{k}_1(\epsilon) +\vec{p}_2^2(0) + 2\vec{p}_2 \cdot \vec{k}(\epsilon) +\vec{k}^2(0) .
    \end{split}
\end{equation}
This piece gives terms only proportional to $\epsilon$, so we can neglect it. On the other hand, by moving the derivative on $\delta'(\tau)$ in $\hat{j}$ to the Green's function, we can find 
\begin{equation}
    \begin{split}
    \hat{j}(\tau)\circ \bar{j}(\tau') &= 
    \int\limits_{-\epsilon}^{\epsilon} d\tau d\tau'\hat{j}(\tau)G(\tau,\tau')\bar{j}(\tau') =  \\
    &= \int \limits_{-\epsilon}^{\epsilon}d\tau d\tau'\left[ \vec{\varepsilon} \delta'(\tau)\right] |\tau - \tau'| \left[- \vec{p}_1 \delta(\tau' + \epsilon) + \vec{p}_2\delta(\tau'- \epsilon) + \vec{k}\delta(\tau')\right] \\
    &= -\int \limits_{-\epsilon}^{\epsilon}d\tau d\tau' \;\vec{\varepsilon} \;\delta(\tau)\; {\rm sign}(\tau-\tau')\;\left[- \vec{p}_1 \delta(\tau' + \epsilon) + \vec{p}_2\delta(\tau'- \epsilon) + \vec{k}\delta(\tau')\right]\\
    &= -\vec{\varepsilon}\cdot \int \limits_{-\epsilon}^{\epsilon} d\tau'\;{\rm sign}(-\tau')\;\left[- \vec{p}_1 \delta(\tau' + \epsilon) + \vec{p}_2\delta(\tau'- \epsilon) + \vec{k}\delta(\tau')\right]\\
    &= -\vec{\varepsilon}\cdot(-\vec{p}_1\;{\rm sign}(\epsilon) + \vec{p}_2\;{\rm sign}(-\epsilon) + \vec{k} \;{\rm sign}(0))\\
    &= \vec{\varepsilon}\cdot(\vec{p}_2 + \vec{p}_1)
    \end{split}
\end{equation}
Similarly, we will get the exact same contribution from $\bar{j}(\tau)\circ \hat{j}(\tau')$. Finally, the last piece, proportional to $\hat{j}\circ\hat{j}$, will be:
\begin{equation}
    \begin{split}
    \hat{j}(\tau)\circ \hat{j}(\tau') &= 
    \int\limits_{-\epsilon}^{\epsilon} d\tau d\tau'\hat{j}(\tau)G(\tau,\tau')\hat{j}(\tau') =  \\
    &= \int \limits_{-\epsilon}^{\epsilon}d\tau d\tau'\; \vec{\varepsilon} \delta'(\tau)\; |\tau - \tau'|  \; \vec{\varepsilon} \delta'(\tau')\; \\
    &= 2\vec{\varepsilon}^2 \int \limits_{-\epsilon}^{\epsilon}d\tau d\tau'\;\delta(\tau)\delta(\tau-\tau')\delta(\tau') = 2\vec{\varepsilon}^2\delta(0),
    \end{split}
\end{equation}
which we set to 0 by the Veltman prescription, following the conventions used in \cite{Schuster:2023jgc}. This leads to the final amplitude:
\begin{equation}
    \begin{split}
        \label{eq: general amplitude momentum app version}
        A &= \mathcal{C}e^{i\xi} \int dt^* e^{-i(E_{in}-E_{out}-\omega)t^*}\int \frac{d^3p_1}{(2\pi)^{d/2}} \frac{d^3p_2}{(2\pi)^{d/2}} \tilde{\psi}_{in}(\vec{p}_1)\tilde{\psi}^*_{out}(\vec{p}_2)\;e^{\frac{\vec{\varepsilon}}{2\tilde{m}}\cdot(\vec{p}_2 + \vec{p}_1)}(2\pi)^3 \delta^{(3)}(P).
    \end{split}
\end{equation}

\section{Ward Identity for Bound States}
\label{app: Ward}
In this section we prove that contributions of $O(\rho^{-1})$ from the vertex operator in Eq. \eqref{eq:CSP_vertex_operator_quotient_form} do not contribute to bound state transition amplitudes. In the full vertex operator, the $O(\rho^{-1})$ terms will come from 
\begin{equation}
\label{eq: 1/rho vertex app verion}
    e^{-ik\cdot z(t)} \ \left(k \cdot \dot{z}(t)/\rho \right).
\end{equation}
Of course, this is a term that differs from the QED vertex by $\epsilon_\mu \rightarrow k_\mu$ and therefore must vanish by the Ward identity in a full theory. To see this, we can write Eq. \eqref{eq: 1/rho vertex app verion} as
\begin{equation}
\label{eq: 1/rho vertex d/dt version}
    e^{ik\cdot z}(k\cdot {\dot{z}}) = -i\del_t e^{ik\cdot z},
\end{equation} 
showing that this term is a total $z_0$ derivative, and reduces to a boundary condition evaluated at the start and end times $t$ and $t'$, which does not contribute to the amplitude.  To prove that claim, we can plug Eq. \eqref{eq: 1/rho vertex d/dt version} into our original amplitude
\begin{equation}
    \begin{split}
        A = -i\int d^3x'\; \psi^*_{out}(\vec{x}')\int d^3x \;\psi_{in}(\vec{x}) \int_{\vec{z}(t)= \vec{x}}^{\vec{z}(t')=\vec{x}'} \mathcal{D}z \; \left(\int_t^{t'} dt^* \frac{d}{dt^*} (e^{i\omega t^* - i\vec{k} \cdot\vec{z}(t^*)})\right) \;e^{i\int_t^{t'} d\tau L_0(\vec{z},\tau)}.
    \end{split}
\end{equation}
This interaction clearly reduces to boundary conditions, yielding a simple path integral that can be computed easily. Carrying out these steps we find 
\begin{equation}
    A = i(e^{i\omega t-iE_{out}(t'-t)} - e^{i\omega t'-iE_{in}(t'-t)} )\int d^3x \; \psi_{in}(\vec{x}) \;e^{- i\vec{k} \cdot \vec{x}} \psi_{out}(\vec{x}).
\end{equation}
Now we can stop and think about what the squared amplitude will look like. Starting with just the first term, which has the only time dependence, we have
\begin{equation}
    \begin{split}
    &\left|\left(e^{i\omega t-iE_{out}(t'-t)} - e^{i\omega t'-iE_{in}(t'-t)} \right)\right|^2 = 2 - 2 \cos((E_{out} + \omega - E_{in})(t'-t))
    \end{split},
\end{equation}
which is of course $0$ if energy is conserved. Therefore, all $O(\rho^{-1})$ contributions to transition amplitudes in bound states vanish.

\section{Non-Relativistic Vertex Operator}
\label{app: non-rel vertex}
\begin{center}
    \begin{table}
    \begin{tabular}
    {|p{1.25cm}|p{3.5cm}|p{3.5cm}|p{3.5cm}|p{3.5cm}|}
     \hline
     & \hfil {$\Delta m = 0$} & \hfil {$\Delta m = 1$} &\hfil {$\Delta m =2$} &\hfil {$\Delta m =3$} \\
     \hline
     &&&&\\
     \hfil $\rho^{-1}$ & \hfil $0$ by Ward  & \hfil $0$ by symmetry   &\hfil $0$ by symmetry  & \hfil $0$ by symmetry \\
     &&&&\\
     \hline
     &&&&\\
     \hfil $\rho^{0}$ & \hfil $0$ by parity & \hfil {\bf Nonzero} $\rightarrow $ QED & \hfil $0$ by symmetry& \hfil $0$ by symmetry\\
     &&&&\\
     \hline
     &&&&\\
     \hfil $\rho^{1}$ & \hfil {\bf Nonzero} & \hfil \vdots & \hfil {\bf Nonzero} & \hfil $0$ by symmetry\\
     &&&&\\
     \hline
     &&&&\\
     \hfil $\rho^{2}$ & \hfil \vdots & \hfil \vdots & \hfil \vdots & \hfil {\bf Nonzero} \\
     &&&&\\
     \hline
    \end{tabular}
    \caption{The lowest order in $\rho$ contributions to a given $\Delta m$ transition are shown. Entries that are zero vanish to all powers in velocity.}
    \label{table: rho perturbation}
    \end{table}
\end{center}
In this section we will justify the non-relativistic expansion of the CSP vertex operator in Sec. \ref{subsec: Vhat CSP}. To do this, we show that for a given transition, the lowest $O(\rho)$ contribution  allowed by parity and symmetry corresponds to the result obtained from the non-relativistic operator in Eq. \eqref{eq: nonrel CSP vertex minimial}. To get a sense for what we mean, consider Table \ref{table: rho perturbation}, showing which matrix elements of the CSP vertex operator are zero, order by order in $\rho$, for different $\Delta m$ transitions (where $m$ is the $\hat{z}$ component of the atom's angular momentum). It is clear that for a given order in $\rho$, the matrix element given by the vertex piece in Eq \eqref{eq: nonrel CSP vertex minimial}:
\begin{equation}
\label{eq: nonrel CSP vertex minimal app version}
    \begin{split}
    \hat{V}^{k,\eta}_{\text{CSP}}(t) &\approx (iq_e)  \;e^{i\omega t}e^{i\frac{\rho}{\omega}\vec{\eta} \cdot \dot{\vec{z}}(t)}\left(\sqrt{2}i\omega/\rho\right),
    \end{split}
\end{equation}
contains the leading order contribution, since all corrections will be higher order in velocity. The task therefore becomes showing that the leading order in $\rho$ result we find by using this non-relativistic operator matches with the first allowed order in $\rho$ contribution to all orders in velocity. 

As a starting point, it is necessary to show that any term that shows up as $\vec{k}\cdot \dot{\vec{z}}$ or $\vec{k}\cdot \vec{z}$ in the vertex operator, will also show up in the same form in the final integral against wave functions. It is for this reason that we need the full, relativistic result in Eq. \eqref{eq: general amp PI solved}:
\begin{equation}
\label{eq: general amp PI solved app version}
    A = \mathcal{C}\; e^{i\xi} \int dt^* e^{-i(E_{in}-E_{out}-\omega)t^*}\; e^{\frac{i\vec{b}\cdot \vec{k}}{2m}} \int d^3x \; \psi_{in}(\vec{x})\psi^*_{out}(\vec{x}-\tfrac{\vec{b}}{m}) e^{-i\vec{k}\cdot \vec{x}}.
\end{equation}
To prove that this equation holds true for the full CSP vertex operator requires only that we can put $\hat{V}_{\rm CSP}^{k,\eta}$ into the exponentiation form of $\hat{V}_{\rm seed}$ as in Eq. \eqref{eq: Vhat QED}. \cite{Schuster:2023jgc} showed that we can do this via the following integration technique:
\begin{equation}
\begin{split}
    \label{eq: b integral full}
&\hat{V}_{\rm CSP}^{k,\eta}(t) = \int [d\mu] e^{ib^\mu(\eta,k)\dot{z}_\mu(t)} e^{ik\cdot z(t)},\\
&[d\mu] \equiv (\sqrt{2} i)^w \frac{d\theta}{\theta^w} d\lambda ds \ e^{-i \lambda \rho}\frac{d}{ds}\left(\frac{2 s}{s^2+\epsilon^2}\right)
\qquad \newvar^\mu \equiv \theta \eta^{\mu} - (s-\lambda\theta) k^{\mu}.
\end{split}
\end{equation}
The integrand of Eq. \eqref{eq: b integral full} matches the form of $\hat{V}_{\rm seed}$ up to a redefinition of $\vec{\varepsilon}$ and an overall constant. Eq. \eqref{eq: general amp PI solved app version} then allows us to recover any interaction term that is polynomial in $\dot{\vec{z}}$ by extracting the appropriate $O(b)$ term.  Setting $\vec{b}$ in this scenario either to $\tfrac{\rho}{\omega} \vec{\eta}$ or $\vec{k}$ lets us analyze interaction terms $\tfrac{\rho}{\omega} \vec{\eta} \cdot \dot{\vec{z}}$ and $\vec{k}\cdot \dot{\vec{z}}$ order by order. One can see that such a Taylor expansion in $b$, means that velocity corrections will show up as $(\vec{k} \cdot \vec{x})^n$ corrections to the final integral (and one can make a similar expansion in the term $e^{-i\vec{k}\cdot \vec{x}}$).

With that understood, we can now analyze Table \ref{table: rho perturbation}. Let us begin by considering any transition for which $\Delta m > [\rho] + 1$ (i.e. the top right of the table). In the same way that any vertex element of the form $(\vec{k}\cdot\dot{\vec{z}})^n$ results in $(\vec{k} \cdot \vec{x})^n$ corrections to the final integral, so too will vertex elements of the form $(\frac{\rho}{\omega}\vec{\eta}\cdot\dot{\vec{z}})^n$ result in $(\frac{\rho}{\omega}\vec{\eta}\cdot\vec{x})^n$ corrections to the final integral. In cylindrical coordinates, each power of $\vec{\eta} \cdot \vec{x}$ will result in one power of $\cos{\tilde{\phi}}$ (where $\tilde{\phi}$ is the angle between $\vec{\eta}$ and $\vec{x}$ in the x-y plane), while $\vec{k}\cdot{\vec{x}}$ will have no $\tilde{\phi}$ dependence for a CSP photon propagating in the $\hat{z}$ direction. The other $\tilde{\phi}$ dependent piece comes from the starting state wavefunction in the form  of $e^{im\tilde{\phi}}$. Therefore the integral over $d\tilde{\phi}$ becomes
\begin{equation}
    \begin{split}
        \int_0^{2\pi}d\tilde{\phi} \cos^n(\tilde{\phi}) e^{im\tilde{\phi}} = \int_0^{2\pi}d\tilde{\phi} \left(\frac{e^{i\tilde{\phi}} + e^{-i\tilde{\phi}}}{2}\right)^n e^{im\tilde{\phi}}.
    \end{split}
\end{equation}
We can now notice that for any integer $m$ and $n$, the only non-zero term will be those for which the exponents cancel. Therefore, for any $n< m$, this integral is zero.  This argument will set the entire top right of the table above the diagonal to zero.  We therefore only have two cells of the table left to analyze, which are the $O(\rho^{-1})$ and $O(\rho^0)$ rows of the $\Delta m = 0$ column. The $O(\rho^{-1})$ term is set to zero by the argument in App. \ref{app: Ward}. In fact, the entire top row of Table \ref{table: rho perturbation} is zero independent of the symmetry argument presented here for the same reason. Finally, we can invoke parity argument to prove that the $O(\rho^0)$, $\Delta m=0$ cell is zero. With no contribution from the wave functions to the $d\tilde{\phi}$ integral, the one power of $\cos{\tilde{\phi}}$ at order $O(\rho^0)$ leads immediately to a vanishing matrix element.  We have therefore shown the first potentially non-zero matrix element for each $\Delta m$ transition order by order in $\rho$. As long as the lowest order in $\rho$ result we find using Eq. \eqref{eq: nonrel CSP vertex minimal app version} matches with the first non-zero elements in Table \ref{table: rho perturbation}, we can ignore any velocity corrections.
\bibliography{ref}

\end{document}